\documentclass[aps, prx, twocolumn, 10pt, superscriptaddress, floatfix, longbibliography, nofootinbib]{revtex4-2}
\usepackage[OT2,T1]{fontenc}

\usepackage{amsmath,amsfonts,amssymb}

\usepackage{bm}
\usepackage{graphicx}

\usepackage{multirow}

\usepackage[a4paper, bindingoffset=0cm, left=2cm, right=2cm, top=2.5cm, bottom=2.5cm, headsep=0.5cm, footskip=1cm]{geometry}

\usepackage{parskip}
\usepackage{tikz-cd}
\DeclareMathAlphabet{\mathscrbf}{OMS}{mdugm}{b}{n}
\usepackage{tikz}
\usetikzlibrary{decorations.markings}

\usepackage{enumerate}
\usepackage[colorlinks, allcolors=blue]{hyperref}

\usepackage{comment}

\usepackage{fancyhdr}
\fancypagestyle{plain}{
  \fancyhf{}
  \fancyfoot[C]{\thepage}
  
}

\usepackage{titlesec}
\usepackage[titles]{tocloft}
\cftsetindents{section}{0em}{1.7em}
\cftsetindents{subsection}{1.7em}{1.2em}
\makeatletter
\renewcommand{\paragraph}{%
  \@startsection{paragraph}{4}%
  {\z@}{1.75ex \@plus 1ex \@minus .2ex}{-1em}%
  {\normalfont\normalsize\bfseries}%
}
\makeatother

\newcommand{\beq}{\begin{equation}}
\newcommand{\eeq}{\end{equation}}
\newcommand{\bem}{\begin{pmatrix}}
\newcommand{\eem}{\end{pmatrix}}

\newcommand{\dd}{\textrm{d}}

\newcommand{\der}{\partial}

\def\diag{\operatorname{diag}}

\def\sgn{\operatorname{sgn}}
\def\Diff{\operatorname{Diff}}

\def\Vect{\operatorname{Vect}}

\newcommand{\epss}{\delta}
\newcommand{\phii}{\varphi}

\newcommand{\ds}{\displaystyle}

\newcommand{\pdag}{^{\vphantom{\dagger}}}

\newcommand{\ppr}{^{\vphantom{\prime}}}

\newcommand{\ie}{\textit{i.e.}\ }
\newcommand{\eg}{\textit{e.g.}\ }
\newcommand{\cf}{\textit{cf.}\ }

\newcommand{\bx}{\mathbf{x}}

\newcommand{\bV}{\mathbf{V}}
\newcommand{\bW}{\mathbf{W}}
\newcommand{\bX}{\mathbf{X}}
\newcommand{\bY}{\mathbf{Y}}

\newcommand{\cF}{{\cal F}}
\newcommand{\cG}{{\cal G}}
\newcommand{\cH}{{\cal H}}

\newcommand{\cM}{{\cal M}}

\newcommand{\cO}{{\cal O}}

\newcommand{\cQ}{{\cal Q}}

\newcommand{\cT}{{\cal T}}
\newcommand{\cU}{\,{\cal U}}

\newcommand{\cu}{\mathfrak{u}}

\newcommand{\sfc}{\mathsf{c}}

\newcommand{\sfC}{\mathsf{C}}

\newcommand{\CC}{\mathbb{C}}

\newcommand{\II}{\mathbb{I}}

\newcommand{\RR}{\mathbb{R}}

\newcommand{\ZZ}{\mathbb{Z}}

\renewcommand\i[1]{\textit{#1}}
\newcommand{\stt}{{\mathcal{T}}}

\begin{document}
%==========================================================

\title{%
Nonequilibrium Probes of Quantum Geometry in Gapless Systems%
}%

\author{Bastien Lapierre}
\email{bastien.lapierre@phys.ens.fr}
\affiliation{Department of Physics, Princeton University, Princeton, New Jersey, 08544, USA\vspace{-1.6mm}}
\affiliation{Philippe Meyer Institute, Physics Department, École Normale Supérieure (ENS), Université PSL, 24 rue Lhomond, F-75231 Paris, France\vspace{-1.6mm}}

\author{Per Moosavi}
\email{per.moosavi@fysik.su.se}
\affiliation{Department of Physics, Stockholm University, 10691 Stockholm, Sweden\vspace{-1.6mm}}
\affiliation{Institute for Theoretical Physics, ETH Zurich, Wolfgang-Pauli-Strasse 27, 8093 Z{\"u}rich, Switzerland\vspace{-1.6mm}}

\author{Blagoje Oblak\vspace{-3mm}}
\email{oblak@math.univ-lyon1.fr}
\affiliation{Université Claude Bernard Lyon 1, ICJ UMR5208, CNRS, 69622 Villeurbanne, France}

\date{February 23, 2026}

\begin{abstract}
Much of our understanding of gapless quantum matter stems from low-energy descriptions using conformal field theory. This is especially true in 1+1 dimensions, where such theories have an infinite-dimensional parameter space induced by their conformal symmetry. We reveal the underlying quantum geometry by considering finite many-body systems driven by time-dependent conformal transformations. For small deformations, perturbation theory predicts absorption rates and linear responses that probe the quantum geometric tensor. For arbitrarily large but adiabatic deformations, we show that periodic drives give rise to nontrivial return amplitudes involving the quantum metric, beyond the familiar leading order that only features a Berry phase. The former is less sensitive to decoherence than the latter, so it can provide robust experimental signatures of our predictions. Our field-theoretic findings are universal, comprising general relations between measurable quantities and quantum geometry that only depend on the emergent effective description. This is supported both by numerical simulations of gapless lattice models, and by exact results for quantum dynamics under certain Floquet drives, probing the full dynamical parameter space.
\end{abstract}

%==========================================================
\maketitle
%==========================================================

\thispagestyle{fancy}
\pagestyle{fancy}

%==========================================================

\setcounter{tocdepth}{2}
\tableofcontents

%==========================================================
\section{Introduction}
\label{Sec:Intro}
%==========================================================

\begin{figure*}[t]
\centering
\includegraphics[width=\textwidth]{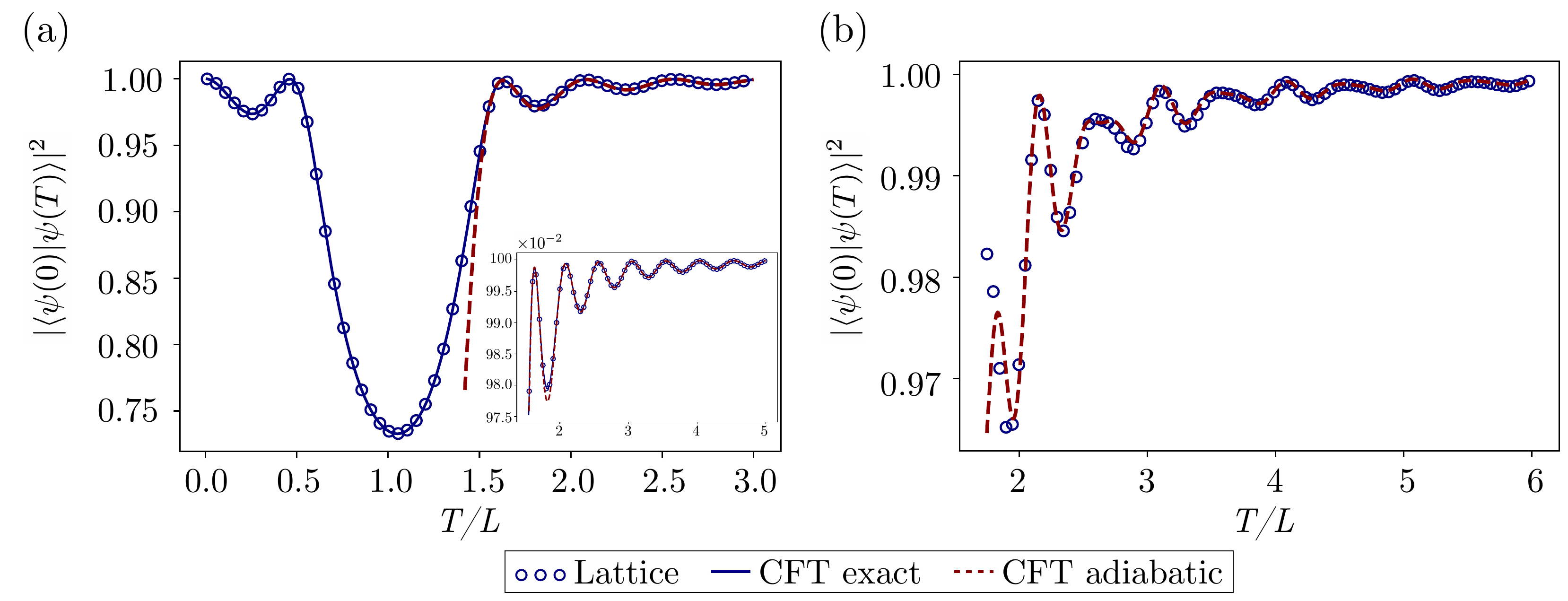}
\caption{%
Return probability $|\langle\psi(0)|\psi(T)\rangle|^2$ as a function of the period $T$, obtained by solving the Schr\"odinger equation for the nonchiral generalization \eqref{fullh} of the driven Hamiltonian \eqref{e11}.
The initial state $|\psi(0)\rangle$ is the ground state of the initial Hamiltonian.
The periodic drives in the two plots are given by (a) the single-harmonic velocity profile $v_t(x) = \cosh(2\lambda) - \sinh(2\lambda) \cos(2\pi k[ x/L - t/T])$ with $\lambda=0.15$ and $k=2$, for which the return probability can be evaluated exactly for any driving period (blue curve, see Sec.~\ref{Sec:SL2R}); (b) the velocity $v_t(x) = v \bigl[ \cosh(2\lambda) - \sinh(2\lambda) \cos(2\pi k[ x/L - t/T])+ \frac{1}{5}\sin(2\pi k'[ x/L - t/T]) \bigr]$ with $\lambda = 0.15$, two harmonics $k=2$, $k'=3$, and a normalization $v$.
In both cases, the return probability goes to $1$ in the adiabatic limit $T/L\to\infty$.
For finite $T/L$, this behavior is corrected by oscillations whose amplitude is essentially the squared quantum distance between two nearby coherent states, which we compute explicitly (dashed red curve, see Sec.~\ref{Sec:QGadiab}).
The blue dots are the corresponding numerical simulations of the spin chain \eqref{eq:H_XXZ_t} illustrated in Fig.~\ref{Fig:SpinChain}, with $\Delta = 0$ and couplings $J_{j}(t)$ of the same form as $v_t(x_j)$ for $N = L = 402$ sites $x_j$ (see Appendix~\ref{App:LattCalcs}).
Note the remarkable agreement between numerics and our analytical CFT results.%
}%
\label{introfig_loschmidt}
\end{figure*}

Half a century ago, a seminal paper by Arnold \cite{Arnold:1966} put forward the idea that fluid flows trace geodesics in an infinite-dimensional manifold, revealing the key role played by infinite-dimensional (metric) geometry in physics \cite{ARS:Kac:1985, KhesinWendt:2009, KhesinEtAl:2021}.
The geometry in such cases is classical, related to the phase space and the kinetic energy of the system.
In recent years though, it has come to be appreciated that \i{quantum} geometry---the geometry of bundles of quantum states over parameter spaces \cite{ProvostVallee:1980nc, AnandanAharonov:1990}---is essential for condensed matter physics \cite{Torma:2023, Yu:2024esf}.
Applications range \eg from band theory \cite{PhysRevB.56.12847, PhysRevLett.82.370, PhysRevB.62.1666} to strongly correlated electrons \cite{PhysRevLett.107.116801, PhysRevB.90.165139, PeottaTorma:2015, PhysRevLett.124.167002, PhysRevB.104.045103, PhysRevB.104.045104, ChenLaw:2024, LiuMeraEtAl:2025}, through quantum phase transitions \cite{VenutiZanardi:2007}, time-dependent variational principles \cite{HacklEtAl:2020, KingEtAl:2025, KramerSaraceno:1981} and Floquet engineering \cite{WeinbergEtAl:2017, schindler2025geometricfloquettheory}.

A natural question is whether Arnold's geometric insight on fluids has a quantum counterpart, and whether it has observable implications in the context of quantum geometry.
That the answer should be positive is supported by
the fact that any many-body quantum system is described, at low energies, by a quantum field theory, whose phase space is necessarily infinite-dimensional.
Coherent states on that phase space define a bundle with an infinite-dimensional quantum geometry, as desired.
For instance, Bloch coherent states for a single spin are labeled by two coordinates on a sphere, but their cousins for spin chains are labeled by infinitely many parameters (two at each lattice site) in the thermodynamic limit.
While infinite-dimensional quantum geometry is ubiquitous in that sense, the nontrivial issue is to find a setup that is observable, suitably universal, and sufficiently tractable even beyond small perturbations in parameter space.

This work aims to provide just such a universal description for gapless quantum matter, supported by both analytical computations and numerics.
Specifically, we consider nonequilibrium setups that probe the quantum metric and the Berry curvature: the former measures the distance between pure states \cite{ProvostVallee:1980nc}, while the latter quantifies phase variations in parameter space \cite{Berry}.
These quantities are what is meant in practice by quantum geometry, and knowing them provides a handle on the dauntingly vast space of quantum states.
In particular, the metric is expected to be less sensitive to decoherence than Berry phases, thereby providing robust experimental signatures to put our predictions to the test.

\begin{figure*}[t]
\centering
\includegraphics[scale=1.4, clip=true, trim=4.5mm -3.4mm 0mm 0mm]{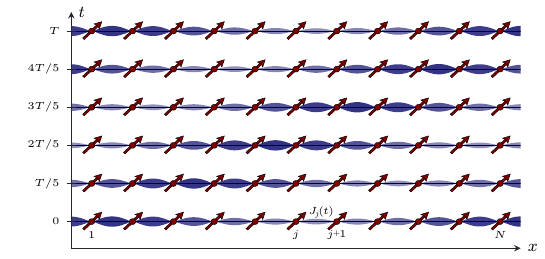}
\caption{%
Cartoon of a periodically driven gapless spin chain \eqref{eq:H_XXZ_t} of length $L \propto N$ with periodic boundary conditions.
The inhomogeneous couplings $J_{j}(t)$ vary on mesoscopic scales in space, and continuously in time with period $T$.%
}%
\label{Fig:SpinChain}
\end{figure*}

%--------------------------------------------------------
\subsection{Driven conformal field theory}
\label{Sec:Intro:dCFT}
%--------------------------------------------------------

Our framework is that of conformal field theory (CFT) in one time and one space dimension (1+1D).
In condensed matter, such models typically emerge as low-energy effective theories of gapless many-body quantum systems \cite{Giamarchi:2003, CCGOR:2011}.
What makes the setup both satisfyingly rich and tractable for our purposes is the emergence of a symmetry under conformal transformations, generated by an infinity of operators---the Fourier modes of the stress-energy tensor---that satisfy the commutation relations of the celebrated Virasoro algebra \cite{BPZ:1984}.
The key idea is to excite the system in a time-dependent manner that switches on operators that are normally silent in CFT, thereby probing directions in quantum parameter space that would not appear for homogeneous systems in equilibrium.

Concretely, consider a right-moving chiral CFT with stress-energy tensor $\stt(x)$, evolving under the time-dependent Hamiltonian\footnote{Units are such that $\hbar = 1$ and velocity is dimensionless, \ie expressed in units of the `speed of light'.}
\beq
H(t) = \int_0^L \dd x\, v_t(x) \stt(x).
\label{e11}
\eeq
Here the velocity profile $v_t(x) > 0$ depends continuously on position $x \in [0, L]$ and time $t \in \RR$, with $L$ the length (volume) of the system.
We impose periodicity in space, namely $\stt(x+L) = \stt(x)$ and $v_t(x+L) = v_t(x)$, and often also in time so that $v_{t+T}(x) = v_t(x)$, with $T$ then the period of the drive.
In standard CFT, the velocity would be homogeneous and constant: $v_t(x) = v = \textrm{const}$.
By contrast, the Hamiltonian \eqref{e11} is deformed to that of a chiral \i{inhomogeneous CFT} \cite{ADSV:2016, DSVC:2017, DSC:2017, GLM:2018, LaMo2:2019, Moosavi:iCFT:2024} with a time-dependent velocity profile $v_t(x)$ generalizing earlier Floquet setups \cite{Wen:2018FCFT, PhysRevX.10.031036, PhysRevResearch.2.023085, LapierreMoosavi:2021, PhysRevResearch.3.023044, 10.21468/SciPostPhys.10.2.049, deBoerEtAl:2023}.
This extra, more general time-dependence will ultimately allow us to probe the quantum geometry of the system, as exemplified in Fig.~\ref{introfig_loschmidt}.

Concentrating on chiral CFTs with spatial periodicity is motivated not only by the simpler presentation, but also by its general applicability.
Indeed, the setting is straightforwardly extended to the nonchiral case of decoupled right and left movers arising in many applications with periodic boundary conditions.
An important example is provided by driven inhomogeneous spin chains, whose low-energy excitations are described by the nonchiral version of \eqref{e11}.
Think \eg of an $XXZ$ spin chain with $N$ sites and Hamiltonian
\beq
\label{eq:H_XXZ_t}
H_{\textrm{XXZ}}(t) = - \sum_{j = 1}^{N} J_{j}(t) \Bigl( S^x_{j} S^x_{j+1} + S^y_{j} S^y_{j+1} - \Delta S^z_{j} S^z_{j+1} \Bigr)
\eeq
with $|\Delta| < 1$ and spin-$1/2$ $\mathfrak{su}(2)$ generators $S^{x,y,z}_{j} \equiv S^{x,y,z}_{j+N}$ at site $j$.
The $J_{j}(t) > 0$ are continuously driven couplings that vary on mesoscopic scales in space\footnote{`Mesoscopic': on length scales much larger than the lattice spacing $L/N$ but much smaller than the system size $L$.} and satisfy $J_{j+N}(t) = J_{j}(t) = J_{j}(t+T)$.
See Fig.~\ref{Fig:SpinChain} for an illustration.
In addition, our setup can describe systems with open boundaries by folding/unfolding techniques adapted to inhomogeneous CFT \cite{GawedzkiKozlowski:2020, LiuEtAl:2025, MCGS:2026}.

Related questions on quantum geometry in CFT have been studied in holography \cite{Oblak:2017,Goto:2021sqx, 10.21468/SciPostPhys.15.5.202, Caputa_2023_coherentstates, Bai:2024azk, Jiang:2024hgt, erdmenger2025driveninhomogeneouscfttheory} and quantum complexity \cite{GiulioTonni:2020, CaputaMagan:2019, ErdmengerEtAl:2025}, following earlier works on Virasoro coadjoint orbits \cite{Kirillov:1987mn, Witten:1987ty, Alekseev:1989, Alekseev:1990mp}.\footnote{There also exists a metric on the space of marginal deformations of 1+1D CFTs \cite{Zamolodchikov:1986gt}, but this is unrelated to our framework since \eqref{e11} is not a marginal deformation of a homogeneous Hamiltonian.}
Models such as \eqref{e11} have also been used to investigate many-body quantum quenches from initial states with inhomogeneous temperature profiles \cite{LLMM2:2017, GLM:2018, Moosavi:iCFT:2024}.
In such cases, the interpretation of $v_t(x)$ as a local (inverse) temperature links thermal transport to metric deformations \cite{Luttinger:1964, GLM:2018, Moosavi:iCFT:2024, BermondEtAl:2024} and provides a smooth version of gluing two systems at different temperatures \cite{BernardDoyon:2012, KarraschIlanMoore:2013, BernardDoyon:2015}.

Note, however, that many of the tools in this paper apply to a wide variety of quantum systems, well beyond CFT.
Indeed, part of our work boils down to unitary representations of continuous groups, and the way these can be used to build time-dependent Hamiltonians.
One instance is the family of Hamiltonians \eqref{e11} for the Virasoro symmetry group of CFT, but many other physical systems fall into the same framework.
Examples include the group $\mathrm{SU}(2)$ describing a spin in a rotating magnetic field \cite{Berry}, or the special linear group $\mathrm{SL}(2,\RR)$ describing both squeezed states in quantum optics \cite{ScullyZubairy:1997} and Hall viscosity \cite{PhysRevLett.75.697,Levay}.
In fact, a close cousin of the chiral theory \eqref{e11} investigated here is the Hamiltonian of edge modes of quantum Hall droplets subjected to time-dependent area-preserving deformations \cite{OLMSE:2024, MOLES:2025, 10.21468/SciPostPhys.15.4.159}.

%--------------------------------------------------------
\subsection{Summary of results}
\label{Sec:Intro:Summary}
%--------------------------------------------------------

\begin{figure}[t]
\centering
\includegraphics[width=.4\textwidth]{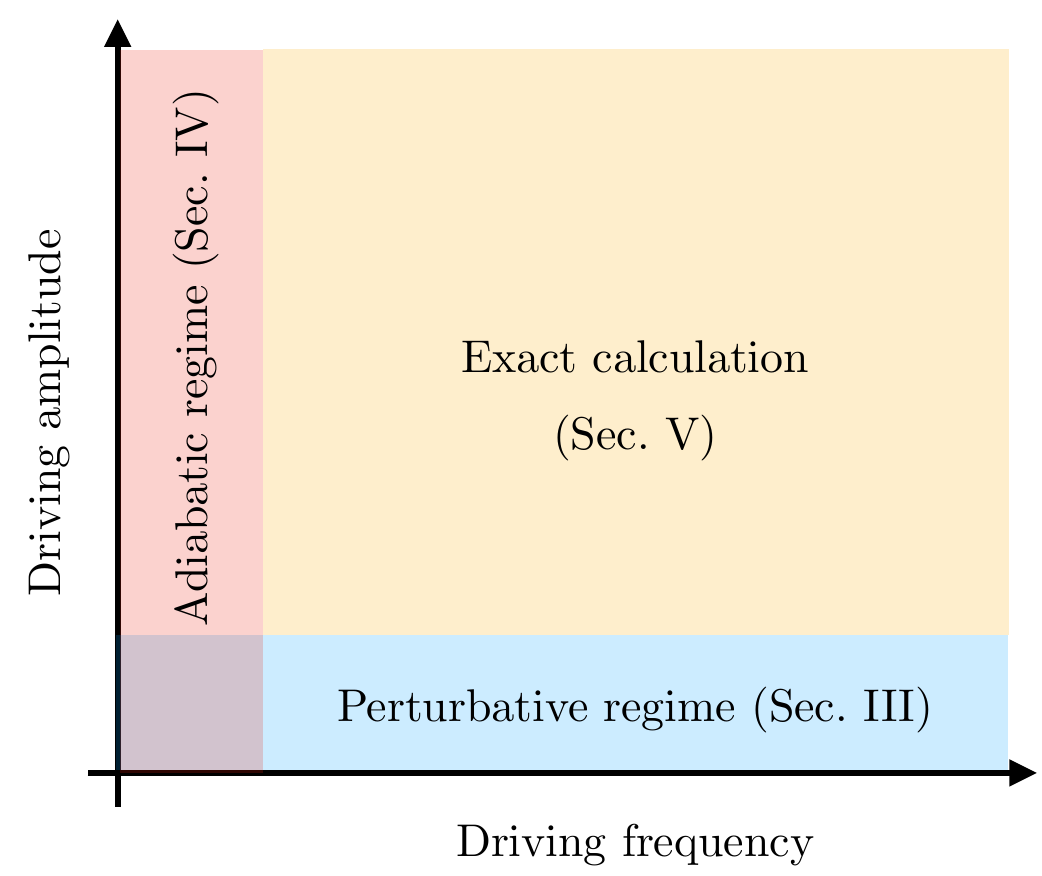}
\caption{%
The different regimes of interest in this work: drives in the perturbative regime are those for which the function $v_t(x)$ in Eq.~\eqref{e11} is close to being uniform, and drives in the adiabatic regime are those for which $v_t(x)$ oscillates slowly with $t$.
When available, exact solutions cover the entire parameter space, but they are limited to specific families of deformations.%
}%
\label{fiReg}
\end{figure}

In studying CFT Hamiltonians of the form \eqref{e11}, we obtain general results split into two categories: \i{perturbative} and \i{adiabatic}; see Fig.~\ref{fiReg}.
These are universal in that they relate quantum geometry to measurable quantities at low energies, in a way that only depends on the central charge of the emergent effective theory.

\paragraph{Perturbative.}
On the perturbative end (Sec.~\ref{Sec:PertQG}), we consider velocities that only deviate slightly from a constant $v_0>0$, namely
\beq
v_t(x)
=
\left\{ \begin{aligned}
& v_0 && \text{for } t<0, \\
& v_0 + \epsilon\,w_t(x) && \text{for } t\geq0,
\end{aligned} \right.
\label{s4}
\eeq
where $\epsilon\ll1$ and $w_t(x) = w_t(x+L)$ is smooth but otherwise arbitrary.
The resulting Hamiltonian \eqref{e11} is close to that of a homogeneous CFT, so familiar formulas on time-dependent perturbation theory apply (provided $L$ is kept finite).
In particular, picking $w_t(x) = (\omega L)^{-1} \cos(\omega t) \partial_x X(x)$ with frequency $\omega > 0$ and a time-independent function $X(x)$ yields an absorption rate $\Gamma(\omega)$ induced by transitions from the CFT ground state to excited states.\footnote{We denote vector fields on the circle as $\bX = X(x)\partial_x$, and similarly for $\bY$ and $\bW$.}
We compute this rate and show, following \cite{OzawaGoldman:2018}, that its integral over frequencies is proportional to the quantum metric.
In formulas,
\beq
\label{eq:abs_rate_intro}
\int_0^{\infty} \dd \omega\,\Gamma(\omega)
\sim
\frac{\epsilon^2}{2L^2} \, \cG(\bX,\bX)
\eeq
to leading order in $\epsilon$, where $\cG(\cdot, \cdot)$ is the quantum metric evaluated at the identity in an infinite-dimensional space of inhomogeneous coherent states.
This link mirrors the fluctuation-dissipation theorem, and is similar in that sense to the defining relation of quantized dichroism for quantum Hall systems \cite{tran2017probing, Tran:2018zvw, Asteria:2018csf, Repellin:2019jes}.
Remarkably, it is possible to give explicit expressions for the quantum metric despite the infinite-dimensional parameter space [see Eqs.~\eqref{cG_XmYn}--\eqref{cG_mn} and \eqref{cG_XY}].

Another natural quantity to investigate is the response of stress-energy expectation values to the perturbation \eqref{s4}, using $w_t(x) = \partial_{x} X_t(x)$ for an arbitrary time-dependent profile $X_t(x)$.
In that context, we pick an arbitrary (smooth) $L$-periodic function $W(x)$, and ask how the time-dependent expectation value of $\cO = \int\dd x\, W(x)\stt(x)$ differs from its equilibrium value.
We show that
\beq
\label{eq:lin_resp_intro}
\langle \cO \rangle(t) - \langle \cO \rangle
\sim
\epsilon \int_{0}^{t} \dd s\, \cF(\partial_s\bW_{t-s},\bX_{s})
\eeq
to leading order in $\epsilon$, where $\bW_t \equiv W(x+t)\partial_x$ and $\cF(\cdot,\cdot)$ is the Berry curvature evaluated at the identity in the same infinite-dimensional parameter space as the metric in Eq.~\eqref{eq:abs_rate_intro}.
Again, the curvature can also be expressed explicitly despite the infinite-dimensional parameter space [see Eqs.~\eqref{cF_XmYn}--\eqref{cF_mn} and \eqref{cF_XY}].

\paragraph{Adiabatic.}
On the adiabatic end (Sec.~\ref{Sec:QGadiab}), we consider Hamiltonians \eqref{e11} whose velocity profile $v_t(x)$ is well away from a constant, but varies slowly in time compared to the system size.
We assume that the system is initially prepared in its ground state $|\psi(0)\rangle$, and investigate its overlap with the state $|\psi(T)\rangle$ at the end of an adiabatic cycle.
To leading order in the adiabatic limit, the only difference between $|\psi(0)\rangle$ and $|\psi(T)\rangle$ is an overall phase---the sum of a dynamical phase and a Berry phase \cite{Oblak:2017}.
But, beyond leading order, there are new subleading corrections in the adiabatic parameter $\epss\sim L/T\ll1$.
Our interest lies in these corrections: they probe the quantum metric of the space of conformal frames---the same metric as in the integrated absorption rate \eqref{eq:abs_rate_intro}.
In formulas, the return probability (Loschmidt echo) is
\beq
\label{introlosch}
\bigl| \langle\psi(0)|\psi(T) \rangle \bigr|^2
\sim
1 - \epss^2 \cG(\bY_T,\bY_T),
\eeq
where $\epss\bY_T$ is a small vector field measuring the deviation between the rays of $|\psi(T)\rangle$ and $|\psi(0)\rangle$.
We compute this deviation exactly for `rotating drives' of the form $v_t(x) = v(x-\omega t)$, with a driving frequency $\omega=2\pi/T$ and a given profile $v(x)$ at $t = 0$.
The resulting return probability \eqref{introlosch} gives rise to the oscillating pattern in Fig.~\ref{introfig_loschmidt}.

It should not come as a surprise that adiabatic overlaps such as \eqref{introlosch} probe the quantum metric.
After all, the latter is the pullback of the Fubini-Study metric that measures, by definition, the overlap of rays in projective Hilbert space.
What is special in our case is that we can compute such overlaps dynamically, at any time $t$, by providing an approximate solution $|\psi(t)\rangle$ of the Schr\"odinger equation for a time-dependent CFT Hamiltonian \eqref{e11}.
More precisely, we compute the time-dependent deviation $\bY_t$ between $|\psi(t)\rangle$ and its leading-order adiabatic approximation, whereupon Eq.~\eqref{introlosch} follows at $t=T$.
A key aspect of this derivation is to identify the `correct' adiabatic parameter $\epss$, thanks to which the adiabatic expansion of $|\psi(t)\rangle$ can in principle be pushed to arbitrarily high order.

\paragraph{Beyond expansions.}
We stress that all the results above are expansions, in different limiting regimes, of the same unified picture; see Fig.~\ref{fiReg}.
However, access to the full range of drives is generally only possible via exact analytical computations for special classes of conformal transformations.
These are studied in Sec.~\ref{Sec:SL2R}, and used there to benchmark our limiting results.
An example is displayed in Fig.~\ref{introfig_loschmidt}(a), showing excellent agreement between exact CFT return probabilities, their adiabatic expansions, and the corresponding numerical lattice computations.

%--------------------------------------------------------
\subsection{Plan of the paper}
\label{Sec:Intro:Plan}
%--------------------------------------------------------

The organization of the paper reflects the thematic split outlined above.
In Sec.~\ref{Sec:VirQG}, we introduce the family of CFT Hamiltonians \eqref{e11} and their quantum geometry.
Sec.~\ref{Sec:PertQG} relates this to observable results, such as transition rates and linear-response coefficients, for small perturbations \eqref{s4}.
Sec.~\ref{Sec:QGadiab} concerns large but adiabatically driven deformations, and the general behavior of the return probability exemplified in Fig.~\ref{introfig_loschmidt} is derived for the important class of rotating drives.
In Sec.~\ref{Sec:SL2R}, we present families of time-dependent deformations that can be solved exactly using finite-dimensional matrices, without resorting to expansions in different limiting regimes.
We conclude in Sec.~\ref{Sec:Disc}, listing some natural extensions and follow-ups.
Some technical details are deferred to appendices.

%==========================================================
\section{Virasoro quantum geometry}
\label{Sec:VirQG}
%==========================================================

This section introduces the quantum geometry of the emergent conformal symmetry of gapless quantum systems, phrased within the context of CFT.
We first recall some basics of conformal transformations and Virasoro group theory, then introduce inhomogeneous CFT and coherent states, and finally turn to our main theme: the quantum geometry of CFT.

%--------------------------------------------------------
\subsection{Virasoro group theory}
\label{Sec:VirQG:VirGT}
%--------------------------------------------------------

We begin by reviewing conformal transformations in any 1+1D CFT.
The presentation starts from the \i{group} of conformal transformations and introduces their \i{algebra} as a second step.\footnote{Due to its infinite dimension, representations of the Virasoro group should strictly speaking be built the other way around: starting from the algebra and integrating to the group \cite{GoodmanWallach:1984, GoodmanWallach:1985}. We deliberately deviate from this, since our presentation generalizes directly to any quantum system with a (finite-dimensional) continuous symmetry group.}
Select formulas are summarized in Appendix~\ref{App:CFTdict}.

\paragraph{Conformal maps.}
Our setup is a 1+1D CFT describing the low-energy dynamics of a gapless many-body quantum system.
For definiteness, consider the theory on a Lorentzian cylinder with time coordinate $t \in \RR$ and spatial coordinate $0 \leq x \leq L$, which we will mostly trade for the angle $\phii\equiv 2\pi x/L$.
The spacetime metric reads $\dd s^2 =-\dd t^2+\dd x^2= -\dd t^2 + (L/2\pi)^2 \dd \phii^2$, with the `speed of light' set to one.
The corresponding conformal group is (up to a zero-mode) a direct product $\Diff(S^1) \times \Diff(S^1)$ \cite{Schottenloher:2008}, where $\Diff(S^1)$ is (the universal cover of) the group of orientation-preserving diffeomorphisms of the circle $S^1$.
Specifically, conformal transformations can be expressed in light-cone coordinates $\phii^\pm = \phii \pm 2\pi t/L$ as
\beq
\label{cot}
\phii^- \mapsto f(\phii^-),
\qquad
\phii^+ \mapsto \bar{f}(\phii^+),
\eeq
where $f$ is any smooth real function that satisfies
\beq
\label{wind}
f'(\phii)>0
\qquad
\text{and}
\qquad
f(\phii+2\pi) = f(\phii) + 2\pi,
\eeq
and similarly for $\bar{f}$.
The group operation in $\Diff(S^1)$ is composition $(f\circ g)(\phii) = f(g(\phii))$, the identity element is $\II(\phii) = \phii$, and $f^{-1}$ denotes the inverse of $f$.

Since right- and left-moving conformal transformations commute, one may restrict attention to a single chirality, \ie deal with $f$ alone (rather than a pair $f,\bar f$).
Time translations and rotations then take the same form $f(\phii) = \phii + \text{const}$, as follows \eg from the definition of the right-moving light-cone coordinate $\phii^-$ in Eq.~\eqref{cot}.
This applies both to chiral systems and to nonchiral systems with periodic boundary conditions.
(For nonchiral systems with \eg Neumann boundary conditions, right and left movers are coupled at the boundaries, but the theory can be unfolded to a chiral one on a cylinder with twice the circumference.) We briefly return to this discussion at the end of Sec.~\ref{Sec:QGadiab}, to explain how right and left chiralities combine in a driven CFT.

Let $\cH$ denote the Hilbert space of the CFT.
In quantum theory, symmetry transformations are typically implemented by unitary operators.
This also holds for conformal transformations $f\in\Diff(S^1)$, which act on $\cH$ through unitary operators $\cU(f)$ \cite{GoodmanWallach:1984, GoodmanWallach:1985}.
These furnish a unitary representation of $\Diff(S^1)$, albeit a projective one.
Indeed,
\beq
\label{UfUg_Vir}
\cU(f) \cU(g)
= e^{i c\hspace{0.7pt}\sfC(f,g)} \cU(f \circ g),
\eeq
where the parameter $c>0$ is the central charge of the CFT, and the functional
\beq
\label{bott}
\sfC(f,g)
\equiv
\oint \frac{\dd \phii}{48\pi}\, \log \bigl[ f'(g(\phii)) \bigr] \frac{g''(\phii)}{g'(\phii)}
\eeq
is known as the Bott cocycle; see \eg \cite{KhesinWendt:2009}.
Here and elsewhere, a prime ${}'$ denotes differentiation with respect to $\phii$ and $\oint \dd \phii$ means
$\int_{0}^{2\pi} \dd \phii$.
Note that $\cU(f^{-1}) = \cU(f)^{-1}$ since $\sfC(f, f^{-1}) = 0$.

\paragraph{Virasoro algebra.}
An infinitesimal conformal map is a function $f(\phii)=\phii+\epsilon X(\phii)$, where $\epsilon\ll1$ and $\bX = X(\phii)\partial_{\phii}$
is a real vector field with $X(\phii+2\pi) = X(\phii)$.
Thus, the Lie algebra of $\Diff(S^1)$ is the space $\Vect(S^1)$ of vector fields on the circle endowed with the standard Lie bracket $[X \partial_\phii, Y \partial_\phii] \equiv ( X Y' - Y X' )\partial_\phii$.\footnote{To find this bracket, take infinitesimal deformations $f(\phii)=\phii+\epsilon X(\phii)$ and $g(\phii) = \phii+\varepsilon Y(\phii)$, and expand their commutator $f\circ g\circ f^{-1} \circ g^{-1}$ in $\epsilon$ and $\varepsilon$.}
We expand them in Fourier modes $\hat{X}_m^{\vphantom{*}} = \hat{X}_{-m}^*$ as
\beq
\bX
=
\sum_{m\in\ZZ}
\hat{X}_{m} e^{i m \phii}\der_\phii.
\label{e29}
\eeq
In quantum theory, infinitesimal conformal transformations are represented by anti-Hermitian operators
\beq
\label{e27}
\cu(\bX)
\equiv
\partial_{\epsilon} \cU(\II+\epsilon\bX) \Big|_{\epsilon=0}
\equiv
-i \oint \dd \phii\, X(\phii) \stt(\phii),
\eeq
where we introduced the Hermitian operator-valued distribution $\stt(\theta) = i \cu(\delta(\phii-\theta)\partial_{\phii})$ in terms of the $2\pi$-periodic delta function $\delta(\phii)$.
We take this $\stt$ to be (the right-moving component of) the stress-energy tensor, as the latter is the generator of conformal transformations.
Expanded in Fourier modes,
\beq
\label{e28}
\stt(\phii)
=
\frac{1}{2\pi} \sum_{n\in\ZZ} e^{in \phii} \left( L_n - \frac{c}{24} \delta_{n,0} \right)
\eeq
with $L_n \equiv i \cu \bigl( e^{-in \phii} \partial_{\phii} \bigr) + \frac{c}{24} \delta_{n,0}$, in terms of which Eqs.~\eqref{e29}--\eqref{e27} yield
\beq
\cu(\bX) = -i \sum_{m} \hat{X}_{m} \left( L_{-m} - \frac{c}{24} \delta_{m,0} \right).
\label{cuX_Lm}
\eeq
The operators $L_{n}$ obtained in this way satisfy $L_{n}^{\dagger} = L_{-n}\pdag$ and the commutation relations
\beq
[L_m,L_n]
= (m-n)L_{m+n}+\frac{c}{12}m(m^2-1)\delta_{m+n,0}
\label{e24}
\eeq
of the Virasoro algebra, consistent with the projective composition law \eqref{UfUg_Vir}; see Appendix~\ref{App:CFTdict} for details.

%--------------------------------------------------------
\subsection{Unitary drives in CFT}
\label{Sec:VirQG:udCFT}
%--------------------------------------------------------

We now turn to the action of conformal transformations and show how this allows one to construct parameter-dependent CFT Hamiltonians, each labeled by a choice of `conformal frame'.

\paragraph{Inhomogeneous CFT.}
In any (chiral) 1+1D CFT, time translations are generated by the zero mode of the stress-energy tensor, so the standard CFT Hamiltonian is
\beq
\label{e211}
H_{0} 
=
\frac{2\pi}{L}\oint \dd\phii\, \stt(\phii)
=
\frac{2\pi}{L} \Bigl( L_{0} - \frac{c}{24} \Bigr).
\eeq
We refer to it as the \i{undeformed} or \i{homogeneous} Hamiltonian, in the sense that it is the Hamiltonian of the theory viewed from a suitable `rest frame'.
Indeed, in CFT, observers are free to perform conformal changes of coordinates \eqref{cot} and measure energy from a different `conformal frame'.
(This is similar to special relativity, where Lorentz boosts affect the value of energy.)
We therefore refer to the Hamiltonian in a different frame labeled by $f\in\Diff(S^1)$, namely
\beq
\label{e212}
H[f]
\equiv
\cU(f) H_{0} \cU(f)^{-1},
\eeq
as the CFT Hamiltonian \i{deformed} by $f$.
This construction forms our starting point:
viewing each operator \eqref{e212} as being labeled by a \i{conformal frame} $f\in\Diff(S^1)$, we will investigate the quantum geometry of the resulting bundle of eigenstates.
The underlying parameter space is a quotient 
\beq
\cM\cong\Diff(S^1)/S^1,
\label{s125q}
\eeq
since the undeformed Hamiltonian \eqref{e211} commutes with $L_0$ but fails to commute with any other Virasoro generator, as is clear from the algebra \eqref{e24}.\footnote{The notation $\Diff(S^1)/S^1$ is standard for the space of all $S^1$ diffeomorphisms modulo rotations (see \eg \cite{Witten:1987ty}). Technically, we should instead write $\Diff(S^1)/\RR$, since our notation `$\Diff(S^1)$' refers to the universal cover of the diffeomorphism group of the circle.}
Topologically, the parameter space \eqref{s125q} is homeomorphic to an (infinite-dimensional) linear space, so it is homotopic to a point; see \textit{e.g.}~\cite{GuieuRoger:2007, Oblak:2016eij}.

As mentioned in Sec.~\ref{Sec:Intro}, our strategy is to probe the geometry of parameter space by considering conformal transformations $f_t \in \Diff(S^1)$ that vary continuously in time $t$.
The result is a CFT governed by a time-dependent deformed Hamiltonian
\beq
\begin{aligned}
H[f_t]
& =
\cU(f_t) H_{0} \cU(f_t)^{-1} \\
& =
\frac{2\pi}{L} \oint \dd \phii\, \cU(f_t) \stt(\phii) \cU(f_t)^{-1},
\end{aligned}
\label{s11}
\eeq
where we used Eq.~\eqref{e211}.
This operator can be made explicit by relying on the usual transformation law of the CFT stress-energy tensor,
\beq
\label{tranT}
\cU(f) \stt(\phii) \cU(f)^{-1}
=
[f'(\phii)]^2 \stt(f(\phii)) - \frac{c}{24\pi} \bigl\{ f, \phii \bigr\},
\eeq
with $\bigl\{ f, \phii \bigr\} \equiv {f'''}/{f'} - \frac{3}{2} \bigl( {f''}/{f'} \bigr)^2$
the Schwarzian derivative of $f$.
Plugging this into Eq.~\eqref{s11} yields 
\beq
H[f_t]
=
\frac{2\pi}{L} \oint
\frac{\dd \phii}{(f_t^{-1})'(\phii)}
\stt(\phii) + C[f_t],
\label{t11}
\eeq
where $C[f]\equiv-\tfrac{c}{12L}\oint\dd\phii\,\{f,\phii\}$ is a real number.
The key point is that Eq.~\eqref{t11} takes the form of a time-dependent inhomogeneous CFT \eqref{e11}, with the (dimensionless) velocity $v_t(x) \equiv V_t(\phii) = 1/(f_t^{-1})'(\phii)$ and the identification $\stt(x) \equiv (2\pi/L)^2 \stt(\phii)$ for $\phii = 2\pi x/L$.
Clearly, such deformations of the homogeneous theory \eqref{e211} leave a huge amount of legroom: by suitably choosing the maps $f_t(\phii)$, one can obtain essentially any profile $v_t(x)$.
The only requirements are $v_t(x)>0$ and the normalization $\frac{1}{L} \int_{0}^{L} \frac{\dd x}{v_t(x)} = \frac{1}{2\pi} \oint \frac{\dd \phii}{V_t(\phii)} = 1$, due to the conditions \eqref{wind}.

\paragraph{Virasoro coherent states.}
The parameter dependence of the Hamiltonian \eqref{e212} entails a similar dependence for its eigenstates.
For definiteness, we focus on coherent states, which are essentially obtained by acting with conformal transformations on a primary state.
The latter is, by definition, a normalized state $|h\rangle$ in $\cH$ labeled by a conformal weight $h>0$ and satisfying
\beq
L_{0} |h\rangle = h|h\rangle,
\quad
L_{m} |h\rangle = 0 \quad \text{for all} \;\, m>0.
\label{e217}
\eeq
Such a state can be obtained from the CFT vacuum $|0\rangle$, which satisfies $L_{m} |0\rangle = 0$ for all $m \geq -1$, by acting on it with the corresponding primary field; see \eg \cite{FMS:1997}.
For our purposes, we may assume that there is at least one primary state in the theory, as our results can equally well be stated for the vacuum by setting $h = 0$.
This is important for our numerical simulations, all of which involve the ground state $|h\rangle=|0\rangle$.

Given a primary state $|h\rangle$, it is standard practice to build a highest-weight representation of the Virasoro algebra \eqref{e24}, analogously to highest-weight representations of simple Lie algebras.\footnote{Technically, the representation is \i{lowest}-weight, but the highest-weight terminology is standard.}
We assume that the representation is unitary, so that $h \geq 0$.
Within that representation, the spectrum of $L_{0}$ is (a subset of) $\{h,h+1,h+2,h+3,\ldots\}$, so $|h\rangle$ is effectively a nondegenerate ground state with a (finite-volume) gap separating it from higher-energy states.
Our bundle of \i{Virasoro coherent states} can then be constructed by acting on $|h\rangle$ with finite conformal transformations:
\beq
\cU(f) |h\rangle
\quad
\text{for} \;\, f \in \Diff(S^1).
\label{e218}
\eeq
Each of these is a normalized, nondegenerate eigenstate of the deformed Hamiltonian \eqref{e212} with energy $(2\pi/L)(h-c/24)$ and a (finite-volume) gap above it.
Such states are coherent in the same sense as Bloch coherent states or squeezed states in quantum optics \cite{Perelomov:1986,ScullyZubairy:1997}, where the group $\Diff(S^1)$ would respectively be replaced by $\mathrm{SU}(2)$ or $\mathrm{SL}(2,\RR)$.

%--------------------------------------------------------
\subsection{Quantum geometric tensor}
\label{Sec:VirQG:QGT}
%--------------------------------------------------------

Having introduced the family of coherent states \eqref{e218}, let us write their quantum geometric tensor.
The latter consists of a metric and a Berry curvature.
We first recall the definition of these concepts in a broader setting, then specialize to the Virasoro case.

\paragraph{Generalities on quantum geometry.}
Consider the Hilbert space $\cH$ of some quantum system.
Given any nonzero state vector $|\psi\rangle\in\cH$, let $[|\psi\rangle] \equiv \{ \lambda|\psi\rangle : \lambda\in\CC \}$ be its ray in $\cH$.
Denote by $\mathbb{P}\cH = (\cH\setminus\{0\})/\CC$ the corresponding complex projective space, \ie the space of rays.
Let the Hamiltonian $H[p]$ depend on some set of continuous parameters that label a point $p$ in some parameter manifold $\cM$.
Finally, let $|\phi(p)\rangle\in\cH$ be a normalized eigenstate of $H[p]$, with a dependence on $p$ that we assume to be continuous.
Then there is a well-defined (smooth) map
\beq
\cM\to\mathbb{P}\cH:
p \mapsto \bigl[|\phi(p)\rangle\bigr],
\label{s17}
\eeq
sending a set $p$ of parameters on the corresponding ray.
This map essentially defines the quantum geometry of $\cM$.
Indeed, $\mathbb{P}\cH$ is a K\"ahler manifold equipped with a Fubini-Study metric and a compatible symplectic form, so their pullback by the map \eqref{s17} yields a (possibly degenerate) \i{quantum metric} $\cG$ and a (possibly degenerate) \i{Berry curvature} $\cF$ on $\cM$.
These pullbacks are respectively given by the (symmetric) real and (antisymmetric) imaginary parts of the quantum geometric tensor $\cQ \equiv \langle\dd\phi|\otimes|\dd\phi\rangle - \langle\dd\phi|\phi\rangle\otimes\langle\phi|\dd\phi\rangle$, where $\dd$ is the exterior derivative on $\cM$ \cite{ProvostVallee:1980nc}.\footnote{The definition of $\cQ$ should be understood as follows: given a point $p\in\cM$ and paths $\gamma(t), \chi(s) \in \cM$ such that $\gamma(0)=\chi(0)=p$, the quantum geometric tensor takes the two tangent vectors $\der_t\gamma|_{t=0}$, $\der_s\chi|_{s=0}$ to return the complex number $\cQ(\der_t\gamma|_{t=0},\der_s\chi|_{s=0}) \equiv \der_s\der_t \big[\langle\phi(\chi(s))|\phi(\gamma(t))\rangle - \langle\phi(\chi(s))|\phi(p)\rangle\langle\phi(p)|\phi(\gamma(t))\rangle\big] |_{t = s = 0}$. This definition is independent of the choice of normalized representatives $|\phi(p)\rangle$ for the rays in the map \eqref{s17}.}
Thus, the metric and Berry curvature read
\begin{subequations}
\label{e220}
\begin{align}
\cG & = \langle\dd\phi|\odot|\dd\phi\rangle - \langle\dd\phi|\phi\rangle\odot\langle\phi|\dd\phi\rangle, \\
\cF & = -2i \langle\dd\phi|\wedge|\dd\phi\rangle,
\end{align}
\end{subequations}
where $T_1 \odot T_2 \equiv(T_1 \otimes T_2 + T_2 \otimes T_1)/2$ and $T_1 \wedge T_2 \equiv(T_1 \otimes T_2 - T_2 \otimes T_1)/2$ respectively denote the symmetrized and antisymmetrized tensor products.
The pair $(\cF,\cG)$ is compatible in the sense of K\"ahler manifolds when the map \eqref{s17} is not only smooth but analytic \cite{Avron:2011ogm}, which will turn out to be the case for the Virasoro coherent states \eqref{e218}.

A special case of interest for us occurs when parameter dependence stems from a unitary group action, as in the deformed Hamiltonian \eqref{e212}.
Specifically, let $G$ be a Lie group (with Lie algebra $\mathfrak{g}$); let $\cU$ be a (possibly projective) unitary representation of $G$ in $\cH$, and let the parameter-dependent Hamiltonian read $H[f] = \cU(f)H_0\cU(f)^{-1}$, where $H_0$ is again some `undeformed' reference Hamiltonian.
This would exactly happen for rotating spins or squeezed states, in which case $G=\mathrm{SU}(2)$ or $G=\mathrm{SL}(2,\RR)$, respectively.
The underlying parameter space is a quotient similar to \eqref{s125q},
\beq
\cM \cong G/G_0,
\label{s34}
\eeq
where $G_0$ is the subgroup of $G$ whose elements leave $H_0$ unchanged.
For both Bloch coherent states and squeezed states, this stabilizer $G_0$ is the group $\mathrm{SO}(2)$ $\cong$ $S^1$, respectively yielding a parameter space that is a sphere $S^2 \cong \mathrm{SU}(2)/S^1$ or a (hyperbolic) plane $\RR^2 \cong \mathrm{SL}(2,\RR)/S^1$.

Now choose some (isolated, nondegenerate) eigenstate $|h\rangle$ of $H_0$, and replace the parameter-dependent eigenstates $|\phi(p)\rangle$ in \eqref{s17} by the states $\cU(f)|h\rangle$, as in Eq.~\eqref{e218}.
The resulting metric and Berry curvature \eqref{e220} are readily found thanks to the unitarity relation $\cU(f)^{-1}=\cU(f)^{\dagger}$, and they can be expressed in terms of the Lie algebra representation defined similarly to Eq.~\eqref{e27}, namely $\cu(\bX)\equiv\der_{\epsilon}\cU(\II+\epsilon\bX)|_{\epsilon=0}$ for $\bX\in\mathfrak{g}$.
When evaluated at the identity coset labeled by $f = \II$ in parameter space \eqref{s34}, the metric and the curvature with arguments $\bX,\bY\in\mathfrak{g}$ read
\begin{subequations}
\label{tb18}
\begin{align}
\label{t18}
\cG(\bX,\bY)
& =
- \frac{1}{2} \langle h| \bigl\{ \cu(\bX),\cu(\bY) \bigr\} |h\rangle \nonumber \\
& \quad\,
    + \langle h| \cu(\bX) |h\rangle \langle h| \cu(\bY) |h\rangle, \\
\label{b18}
\cF(\bX,\bY)
& = 
i \langle h| \bigl[ \cu(\bX),\cu(\bY) \bigr] |h\rangle,
\end{align}
\end{subequations}
where $\{\cdot,\cdot\}$ and $[\cdot,\cdot]$ respectively denote the anticommutator and the commutator of operators in $\cH$.\footnote{We are abusing notation: the arguments of the tensors in Eqs.~\eqref{tb18}--\eqref{e226} should be tangent vectors of the manifold \eqref{s34}, not Lie algebra elements. The reason this is harmless is because both quantities in Eq.~\eqref{tb18} vanish when either $\bX$ or $\bY$ belong to the Lie algebra of the stabilizer $G_0$, and because $\bX$ and $\bY$ define unique tangent vectors of the quotient $G/G_0$.}
In particular, the squared quantum norm of $\bX$ at the identity is
\beq
\cG(\bX,\bX)
= - \langle h| \cu(\bX)^2 |h\rangle
    + \langle h| \cu(\bX) |h\rangle^2,
\label{e226}
\eeq
which is nothing but the variance of the Hermitian operator $i\cu(\bX)$.
Note that both the metric and the curvature are generally degenerate on the manifold \eqref{s34}, though they can be nondegenerate on further quotient spaces thereof: we will see examples of this below.
(No such complications arise for Bloch coherent states or squeezed states.)

The quantum metric and the Berry curvature can similarly be evaluated at any other point $fG_0$ in parameter space \eqref{s34}.
In fact, they are fully determined by their expressions \eqref{tb18} at the identity.
The only difference is that the Lie algebra elements $\bX$ and $\bY$ are then replaced by images of the Maurer-Cartan form acting on tangent vectors at $f$.
This is to say that the quantum geometric tensor is invariant under $G$, which stems here from the form $\cU(f)H_0\cU(f)^{-1}$ of the deformed Hamiltonian.
See Appendix~\ref{App:VirQGnonId} for further details, which we omit here since they will seldom be needed below.

\paragraph{Virasoro case.}
The formulas in Eq.~\eqref{tb18} readily apply to the Virasoro coherent states of Sec.~\ref{Sec:VirQG:udCFT}, for which $G = \Diff(S^1)$ and $G_0 = S^1$.
In practice, the only complication is that $G$ is now an infinite-dimensional group, so any tensor field on $G$ is a functional of a function $f(\phii)$.
What simplifies matters is that these tensor fields are invariant under $\Diff(S^1)$, hence wholly determined by their expression at the identity $\II(\phii)=\phii$.
For this reason, we now describe the quantum geometric tensor of CFTs by focusing entirely on Eq.~\eqref{tb18} at the identity.
Minimal effort is then needed to extend the result to fully fledged tensor fields on the entire group manifold, but this last bit is less essential for us and is relegated to Appendix~\ref{App:VirQGnonId} for brevity.
Explicit expressions for the symplectic form and the metric on the parameter space \eqref{s125q} have long been known in the context of Virasoro orbits \cite{Kirillov:1987mn}.
In particular, the Berry curvature $\cF$ can be read off from \cite{Oblak:2017} and coincides with the symplectic form written \eg in \cite{Witten:1987ty, Alekseev:1989, Alekseev:1990mp}.
To our knowledge, however, the application of these objects to driven quantum many-body systems is new.

Consider first the ingredients needed in Eq.~\eqref{tb18}: $\bX,\bY$ are vector fields on the circle, the Lie algebra representation $\cu$ is the one in Eq.~\eqref{e27}, and $|h\rangle$ is the highest-weight state introduced in Eq.~\eqref{e217}.
Fourier expanding $\cu(\bX),\cu(\bY)$ as in Eq.~\eqref{e28} and using the highest-weight conditions \eqref{e217} then yields
\begin{subequations}
\label{cG_cF_XmYn}
\begin{align}
\cG(\bX,\bY)
& =
\sum_{m\in\ZZ} |m|\Big(h + \frac{c}{24} (m^2-1)\Big) \hat{X}_{-m} \hat{Y}_{m} \nonumber \\
& \equiv
\sum_{m,n\in\ZZ} \cG_{m,n} \hat{X}_{m}\hat{Y}_{n},
\label{cG_XmYn} \\
\cF(\bX,\bY)
& =
-2i\sum_{m\in\ZZ}m \Big(h + \frac{c}{24} (m^2-1)\Big)\hat{X}_{-m} \hat{Y}_{m} \nonumber \\
& \equiv
\sum_{m,n\in\ZZ} \cF_{m,n} \hat{X}_{m}\hat{Y}_{n},
\label{cF_XmYn}
\end{align}
\end{subequations}
where the respective components of the metric and the Berry curvature are
\begin{subequations}
\label{cG_cF_mn}
\begin{align}
\cG_{m,n}
& \equiv
\frac{1}{2} \langle h| \bigl\{ L_{-m}, L_{-n} \bigr\} |h\rangle - \langle h| L_{-m} |h\rangle \langle h| L_{-n} |h\rangle \nonumber \\
& =
|m| \Big(h + \frac{c}{24} (m^2-1)\Big) \delta_{m+n,0}, \label{cG_mn} \\
\cF_{m,n}
& \equiv
-i\langle h|[L_{-m}, L_{-n}]|h\rangle \nonumber \\
& =
2im \Big(h + \frac{c}{24} (m^2-1)\Big) \delta_{m+n,0}. \label{cF_mn}
\end{align}
\end{subequations}
That there are infinitely many components reflects the fact that parameter space is infinite-dimensional.
The components with $m=0$ or $n=0$ vanish, because parameter space is a quotient \eqref{s125q}.
Both the metric and the curvature are nondegenerate on the manifold \eqref{s125q} if $h>0$; by contrast, if $h=0$, the directions $m=\pm1$ become degenerate due to the vanishing factor $m^2-1$ in Eq.~\eqref{cG_cF_mn}.\footnote{\label{foom}The cancellation stems from the fact that the CFT vacuum $|0\rangle$ defined below Eq.~\eqref{e217} is invariant under the $\mathrm{SL}(2,\RR)$ group generated by $L_{-1}$, $L_0$, $L_1$, which is not true for a generic highest-weight state $|h\rangle$. We return to this in Sec.~\ref{Sec:SL2R}.}
This is ultimately because the quantum geometry of coherent states \eqref{e218} is that of a coadjoint orbit \cite{Boya}---in the case at hand, an orbit of the Virasoro group \cite{Lazutkin, Kirillov, Witten:1987ty, Balog:1997zz}.
Finally, that the components \eqref{cG_cF_mn} are identical up to an absolute value of $m\in\ZZ$ in the metric (and a trivial rescaling) reflects the fact that the quantum geometry on the manifold \eqref{s125q} is K\"ahler, as is indeed the case for Virasoro coadjoint orbits \cite{Kirillov:1987mn, Witten:1987ty, KhesinWendt:2009}.
One specifically has $\cG(\bX,\bY) = \tfrac{1}{2}\cF(\bX,J\bY)$, where the complex structure $J$ maps each Fourier mode $\hat Y_n$ on $-i\sgn(n) \hat Y_n$.
This K\"ahler property is also true for Bloch coherent states and squeezed states, of which our current setup is an infinite-dimensional analogue.

While it is convenient to write the metric and Berry curvature in terms of Fourier modes as in Eq.~\eqref{cG_cF_XmYn}, it is also possible to express them as functionals of local vector fields.
For the curvature, this can be done \eg by inverse Fourier transforming Eq.~\eqref{cF_XmYn}, which yields
\beq
\cF(\bX,\bY)
= -2 \oint\frac{\dd\phii}{2\pi}
  \Bigl[ \Bigl(h-\frac{c}{24}\Bigr)XY' - \frac{c}{24}XY''' \Bigr].
\label{cF_XY}
\eeq
The analogous computation for the quantum metric is more involved, as the metric turns out to be a nonlocal functional of $\bX,\bY$.
This is due to the absolute value of $m$ in Eq.~\eqref{cG_mn}, which gives rise to the propagator $\sum_{m\in\ZZ}|m|e^{im(\theta-\phii)}=-[2\sin^2([\theta-\phii]/2)]^{-1}$.
Thus, one eventually finds
\begin{align}
\label{cG_XY}
\cG(\bX,\bY)
& =
\oint \frac{\dd\phii\,\dd\theta}{16\pi^2 \sin^2([\phii - \theta]/2)} \\
& \quad \times \biggl\{ \Bigl(h-\frac{c}{24}\Bigr) [X(\phii){-}X(\theta)][Y(\phii){-}Y(\theta)] \nonumber \\
& \quad \qquad + \frac{c}{24}[X'(\phii){-}X'(\theta)][Y'(\phii){-}Y'(\theta)] \biggr\} \nonumber
\end{align}
for the Virasoro quantum metric at the identity.
Note how different this is from the Berry curvature \eqref{cF_XY}, all because of the seemingly innocuous distinction between \eqref{cG_mn} and \eqref{cF_mn}.
In this sense, the K\"ahler structure of the parameter space \eqref{s125q} is manifest in Fourier space, but not in position space.

%==========================================================
\section{Perturbative quantum geometry}
\label{Sec:PertQG}
%==========================================================

The simplest probes of quantum geometry for the coherent states \eqref{e218} are provided by \i{perturbative} drives, \ie deformed Hamiltonians \eqref{s11} whose $f_t$ is infinitesimally close to the identity.
Such setups cannot probe global properties (which involve finite $f_t$), but they suffice to determine the quantum geometric tensor at a single point in parameter space, namely that of the undeformed Hamiltonian \eqref{e211}.
Then, the key simplification is that the Lie algebra \eqref{e24} determines everything, without involving finite conformal transformations.
Homogeneity of the parameter space \eqref{s125q} further means that quantum geometry at the identity determines quantum geometry at any other point, so there is no loss of generality to focus on infinitesimal drives as far as \i{local} quantum geometry is concerned.

In this section, we consider such perturbative drives to derive two universal relations between (i) the quantum metric and absorption rates, and (ii) the Berry curvature and linear responses.

%--------------------------------------------------------
\subsection{Quantum metric from absorption rates}
\label{Sec:PertQG:Rates}
%--------------------------------------------------------

Keeping in mind the fluctuation-dissipation theorem, the expression \eqref{e226} for the squared quantum norm as a variance suggests that the quantum metric can be probed in admittance experiments \cite{tran2017probing,Tran:2018zvw,Asteria:2018csf,Repellin:2019jes}.
Such a proposal was indeed put forward in \cite{OzawaGoldman:2018}.
Here, we develop these ideas for the infinite-dimensional quantum geometry of 1+1D CFT.

\paragraph{Transition rates and perturbative drives.}
Consider a time-dependent Hamiltonian \eqref{s11} whose deformation $f_t(\phii) = \phii + \epsilon X_t(\phii)$ is close to the identity in the sense that $\epsilon$ is small, while $\bX_t$ is some unspecified time-dependent vector field.
Then, to first order in $\epsilon$, the perturbed Hamiltonian reads
\beq
H(t)
= H_{0} + \epsilon \bigl[ \cu(\bX_t), H_{0} \bigr] + O(\epsilon^2),
\label{Ht_uXt}
\eeq
where $H_{0}$ is the homogeneous Hamiltonian \eqref{e211} and $\cu(\cdot)$ is the algebra representation \eqref{e27}.
We assume throughout that $X_t(\phii)$ vanishes for $t<0$, \ie the perturbation is `switched on' at $t=0$.
In the present case, we further assume that the perturbation for $t > 0$ is harmonic with frequency $\omega>0$ so that
\beq
X_t(\phii)
=
\left\{
\begin{aligned}
& 0 && \text{for} \;\,  t < 0, \vspace{1.5mm} \\
& \ds \frac{1}{\omega L}\cos(\omega t) X(\phii) && \text{for} \;\, t>0,
\end{aligned}
\right.
\label{ss15}
\eeq
where $X(\phii)$ is some dimensionless, time-independent profile.
Note that the perturbation scales as $1/\omega$: the higher the frequency, the weaker the perturbation.
This will turn out to be crucial for the link with the quantum metric \cite{OzawaGoldman:2018}.

The Hamiltonian \eqref{Ht_uXt} for $\bX_t$ given by Eq.~\eqref{ss15} falls into the usual regime of time-dependent perturbation theory.
In particular, one can perturbatively compute late-time transition rates between eigenstates of $H_0$.
We focus here on transitions from the highest-weight state $|h\rangle$ of Eq.~\eqref{e217} to any other eigenstate, say $|\phi\rangle$, with an energy difference $E_{\phi}>0$ above that of $|h\rangle$.
To leading order in $\epsilon$, the resulting transition rate is
\beq
\begin{aligned}
\Gamma_{\phi}(\omega)
& \sim
\frac{\epsilon^2}{2\omega^2L^2} \Bigl| \big\langle\phi\big| \bigl[ \cu(\bX), H_0 \bigr] \big|h\big\rangle \Bigr|^2 \delta\big(\omega-E_{\phi}\big) \\
& =
\frac{\epsilon^2}{2L^2} \bigl|\langle\phi|\cu(\bX)|h\rangle\bigr|^2 \delta\big(\omega-E_{\phi}\big).
\end{aligned}
\label{t15}
\eeq
The first formula here is standard in quantum theory.
The second step follows from the $H_0$ in the commutator and the fact that $|h\rangle$ and $|\phi\rangle$ are its eigenstates, so that the difference $E_{\phi}$ of their energies cancels against the factor $1/\omega^2$ due to the delta function.

\paragraph{Integrated absorption rates.}
Eq.~\eqref{t15} can be used to obtain the absorption rate induced by the perturbation \eqref{Ht_uXt} by summing over all possible final states of the transition.
Since the deformed Hamiltonian \eqref{Ht_uXt} is an infinitesimal form of unitary conjugation \eqref{e212}, the only nonzero transition amplitudes are those whose final state $|\phi\rangle$ is a descendant state of $|h\rangle$, \ie one that belongs to the space of the representation based on $|h\rangle$.
The absorption rate thus reads
\beq
\Gamma(\omega)
\sim
\frac{\epsilon^2}{2L^2}
\sum_{k=1}^{\infty} \bigl| \langle\phi_k| \cu(\bX) |h\rangle \bigr|^2 \delta(\omega-E_{\phi_k}),
\label{s16}
\eeq
where we label each descendant $|\phi_k\rangle$ of $|h\rangle$ by an integer $k = 1, 2, \ldots$.
How exactly one chooses the labeling has no importance: the only key points are (i) that the spectrum of $H_0$ is discrete thanks to the finite size $L$, and (ii) that $|h\rangle$ is a nondegenerate eigenstate of $H_0$ within its irreducible representation.

Integrating the rate \eqref{s16} over all frequencies yields
\beq
\begin{aligned}
\int_0^{\infty}\dd\omega\,\Gamma(\omega)
& \sim
\frac{\epsilon^2}{2L^2}  \sum_{k=1}^{\infty} \bigl| \langle\phi_k| \cu(\bX) |h\rangle \bigr|^2 \\
& = 
\frac{\epsilon^2}{2L^2}  \Bigl( \langle h| \cu(\bX) |h\rangle^2 - \langle h| \cu(\bX)^2 |h\rangle \Bigr),
\end{aligned}
\label{t17}
\eeq
upon using that $\cu(\cdot)$ is anti-Hermitian and writing the identity operator as $\sum_{k=1}^{\infty} |\phi_k\rangle \langle\phi_k| + |h\rangle \langle h|$ within the irreducible representation of $|h\rangle$.
The result \eqref{t17} manifestly involves the variance of $i\cu(\bX)$, which we found in Eq.~\eqref{e226} to be the squared quantum norm of $\bX$.
In other words,
\beq
\int_0^{\infty}\dd\omega\,\Gamma(\omega)
\sim
\frac{\epsilon^2}{2L^2}\,\cG(\bX,\bX),
\label{s19}
\eeq
where $\cG$ is the quantum metric at the identity coset in the infinite-dimensional parameter space \eqref{s125q}.
This is the result announced in Eq.~\eqref{eq:abs_rate_intro}.
It is consistent with the protocol proposed earlier in \cite{OzawaGoldman:2018}, applied there to two-level systems and Bloch bands.
As in \cite{OzawaGoldman:2018}, the same protocol can be used to obtain any value $\cG(\bX,\bY)$, since $\cG(\bX,\bY) = [\cG(\bX+\bY,\bX+\bY) - \cG(\bX,\bX) - \cG(\bY,\bY)]/2$.

%--------------------------------------------------------
\subsection{Berry curvature from linear response}
\label{Sec:PertQG:LinResp}
%--------------------------------------------------------

Let us now link the Berry curvature \eqref{b18} to the linear response of the stress-energy tensor to perturbations of the form \eqref{Ht_uXt}.
This serves several purposes: it highlights the K\"ahler structure of the parameter space \eqref{s125q}, it reaffirms the ballistic transport properties of CFT \cite{GLM:2018} in the language of viscosity, and it lays the foundations for the adiabatic considerations of Sec.~\ref{Sec:QGadiab}, where Virasoro Berry phases \cite{Oblak:2017} will make an appearance.
We will also briefly return to linear and nonlinear responses in Sec.~\ref{Sec:SL2R}.

\paragraph{Linear response to Virasoro drives.}
Consider again a time-dependent Hamiltonian $H(t)$ of the form \eqref{Ht_uXt}, for a system prepared in the eigenstate $|h\rangle$ of $H_{0}$ at $t = 0$.
In contrast to Sec.~\ref{Sec:PertQG:Rates}, let the vector field $\bX_t=X_t(\phii)\der_\phii$ have an arbitrary time-dependence, with modes $\hat{X}_{m}(t)$ in the Fourier decomposition \eqref{e29}.
We are interested in the ensuing response of expectation values of Hermitian operators $i\cu(\bW)$ generated by the stress-energy tensor \eqref{e27} for any vector field $\bW = W(\phii)\partial_\phii$.
Denote by $|\phi(t)\rangle$ the state at time $t$ governed by the full time-dependent Hamiltonian.
Then, standard linear-response arguments in the spirit of \cite{Kubo:1957mj} yield (see Appendix~\ref{App:LinResp})
\begin{align}
\label{htcOht_suscep}
& \langle\phi(t)|i\cu(\bW)|\phi(t)\rangle
= \hat W_0 \Bigl(h-\frac{c}{24}\Bigr) \\
& \qquad + \epsilon \sum_{m,n \neq 0} \hat{W}_{m}
  \int_{0}^{t} \dd s\, \chi_{m,n}(t,s)\, \hat{X}_{n}(s)
  + O(\epsilon^2) \nonumber 
\end{align}
with the susceptibility
\beq
\label{chi_mn}
\begin{aligned}
\chi_{m,n}(t,s)
& \equiv
\partial_{s} \cF_{m,n} e^{2\pi i(mt + ns)/L} \\
& = 
- i \frac{2\pi}{L} m \cF_{m,-m} e^{2\pi im(t-s)/L} \delta_{m+n,0}
\end{aligned}
\eeq
in terms of the Berry curvature \eqref{cF_XmYn}--\eqref{cF_mn} in Fourier space.
This is the result announced in Eq.~\eqref{eq:lin_resp_intro}, written here in Fourier modes, but expressible in real space using Eq.~\eqref{cF_XY}.
Formally picking $\bW = \delta(\phii-\theta)\der_\phii$, \ie the modes $\hat{W}_m = e^{-im\theta}/{2\pi}$, one immediately finds
\begin{align}
\label{s145}
& \langle\phi(t)|\stt(\theta)|\phi(t)\rangle\\
& =
\frac{1}{2\pi}\Bigl(h-\frac{c}{24}\Bigr)
- 2\epsilon \int_0^t \frac{\dd s}{L}
\Bigl[
\Bigl(h-\frac{c}{24}\Bigr)\der_\theta^2
- \frac{c}{24}\der_\theta^4
\Bigr] \nonumber \\
& \qquad\qquad\qquad\qquad\quad \times
X_s\Bigl(\theta - 2\pi[t-s]/L\Bigr)
+ O(\epsilon^2) \nonumber 
\end{align}
for the stress-energy tensor.
This amounts to rewriting Eqs.~\eqref{htcOht_suscep}--\eqref{chi_mn} in terms of the real-space, functional Berry curvature \eqref{cF_XY} rather than in Fourier space.
Note that the zero mode of $X_t(\phii)$ does not contribute, as was to be expected on the parameter space \eqref{s125q}.
Lastly, while we restricted ourselves here to linear response, one may similarly compute nonlinear-response terms, which probe higher-order quantum-geometric quantities \cite{HetenyiLevay:2023}.

\paragraph{Linear response in the adiabatic regime.}
The perturbation $X_t(\phii)$ is `adiabatic' when each Fourier mode $\hat{X}_m(t)$ (with $m\neq0$ without loss of generality) satisfies $L|\der_t^{k+1}\hat{X}_m(t)| \ll |\der_t^{k}\hat{X}_m(t)|$ for any $k\geq1$.
In this regime, it is meaningful to expand expectation values in time derivatives of the perturbation.
Eq.~\eqref{htcOht_suscep} then becomes (again, see Appendix~\ref{App:LinResp})
\begin{align}
\label{htcOht_grad_exp}
& \langle\phi(t)| i\cu(\bW) |\phi(t)\rangle
\sim
\hat{W}_{0} \Bigl( h - \frac{c}{24} \Bigr)
+ \epsilon \sum_{m \neq 0} \cF_{m,-m} \nonumber \\
& \times \hat{W}_{m}
\biggl(
1 + \frac{L\der_t}{2\pi i m}
+ O\bigl(L^2\der_t^2/m^2\bigr)
\biggr)
\hat{X}_{-m}(t)
\end{align}
to first order in $\epsilon$, again involving $\cF_{m,n}$ in Eq.~\eqref{cF_mn}.
The adiabatic version of Eq.~\eqref{s145} is
\begin{align}
\label{e240}
& \langle\phi(t)| \stt(\phii) |\phi(t)\rangle
\sim
\frac{1}{2\pi} \Bigl( h - \frac{c}{24} \Bigr) \\
& \qquad\quad
- \frac{\epsilon}{\pi}
\Bigl(
h - \frac{c}{24}
- \frac{c}{24}\partial_{-}^2
\Bigr)
\partial_{-}
\bigl[
X_t(\phii)-\hat{X}_0(t)
\bigr] \nonumber
\end{align}
in terms of the partial derivative $\der_-\equiv\der_\phii-\tfrac{L}{2\pi}\der_t$ along the light-cone coordinate $\phii^-$ defined above Eq.~\eqref{cot}.

The links to quantum geometry are manifest, as all formulas since Eq.~\eqref{htcOht_suscep} involve the Berry curvature \eqref{cF_mn} or its functional analogue \eqref{cF_XY}.
In the case at hand, Berry curvature and Virasoro algebra are so closely related that one can nearly read off the Virasoro commutation relations from the linear response, \eg through Eq.~\eqref{e240} with the ubiquitous higher-derivative contribution proportional to the central charge $c$.
(Similar formulas were previously obtained in \cite{Moosavi:iCFT:2024} for inhomogeneous CFT out of equilibrium.)
Note also the similarity between Eqs.~\eqref{htcOht_grad_exp}--\eqref{e240} and the standard definition of viscosity, where the stress-energy expectation value depends on the rate of strain.
Indeed, the odd viscosity of quantum Hall systems is precisely their linear response to adiabatic deformations of the metric \cite{PhysRevLett.75.697}.
Such deformations are implemented by unitary operators \cite{Levay,10.21468/SciPostPhys.15.4.159}, similar to our Eq.~\eqref{e212} for CFTs.
We stress, however, that the viscosity tensor vanishes in standard 1+1D CFT: its odd part is trivially zero for dimensional reasons, and its dissipative part vanishes since all perturbations propagate ballistically.
Instead, our result \eqref{e240} should be seen as a response of the thermal current to perturbations of the metric, \ie a ballistic contribution to the thermal conductance of a 1+1D CFT \cite{Moosavi:iCFT:2024, GLM:2018}.

%==========================================================
\section{Adiabatic quantum geometry}
\label{Sec:QGadiab}
%==========================================================

Having investigated perturbative probes of Virasoro quantum geometry, we now turn to finite deformations $f_t \in \Diff(S^1)$ in the Hamiltonian \eqref{s11}.
We assume throughout that these deformations are adiabatic and periodic, with period $T$ so that $H[f_T] = H[f_0]$.\footnote{Beware that this does not imply $f_T=f_0$, but only $f_T(\phii)=f_0(\phii+\theta)$ for some angle $\theta$.}
Then the final state vector $|\psi(T)\rangle$ is nearly proportional to the initial one, $|\psi(0)\rangle$, and one is typically interested in their overlap,
\beq
\langle\psi(0)|\psi(T)\rangle.
\label{s165}
\eeq
To leading order in the adiabatic limit, the overlap has unit complex norm and involves a Berry phase $\gamma_{\textrm{B}}$ given by the integral of the Berry connection along the curve $f_t$.
This was studied in detail in \cite{Oblak:2017}, so we omit the derivation and merely state the result for later reference: given a drive $f_t$, the Virasoro Berry phase is\footnote{The conventions here differ from \cite{Oblak:2017} by an overall sign.}
\begin{align}
\gamma_{\textrm{B}}[f_t]
& =
\int_0^T \!\!\! \dd t
\oint\frac{\dd\phii}{2\pi}
\frac{\der_tf_t(\phii)}{f_t'(\phii)}
\Biggl[h-\frac{c}{24}
+\frac{c}{24} \biggl(\!\frac{f_t''(\phii)}{f_t'(\phii)}\!\biggr)^{\!\!\prime\,}
\Biggr] \nonumber \\
& \quad
- \Bigl( h - \frac{c}{24} \Bigr)\, f_0^{-1}(f_T(0)),
\label{s43}
\end{align}
where $h$ is the highest weight introduced in \eqref{e217} and $c$ is the central charge.
Instead, our goal here is to go beyond leading order and capture the slight misalignment between initial and final states, resulting in a nontrivial return probability given by the squared complex norm of \eqref{s165}.

Since the deformations in Eq.~\eqref{e212} are implemented by unitary operators in a representation of the group $\Diff(S^1)$, the return amplitude \eqref{s165} is intimately linked to group theory.
This is also true of rotating spins and squeezed states [with groups $\mathrm{SU}(2)$ and $\mathrm{SL}(2,\RR)$, respectively], indeed of all generalized coherent states whose parameter space is a quotient \eqref{s34} \cite{Perelomov:1986, ScullyZubairy:1997}.
What follows will thus be phrased, partly, in terms of group manifolds.
While we focus for definiteness on $\Diff(S^1)$, most of our arguments carry over to other groups and coherent states previously studied elsewhere, including \eg \cite{AnandanAharonov:1990, BraunlichGrafOrtelli:2010, AvronEtAl:2012}.

The plan is as follows.
First, we show in general terms how the Schr\"odinger equation with a time-dependent Hamiltonian \eqref{e212} can be solved in terms of paths in the $\Diff(S^1)$ group manifold.
This expresses the return amplitude \eqref{s165} as an expectation value in a highest-weight state, which ultimately makes it manifest that the return probability is related to the quantum metric \eqref{e226}.
Second, we investigate the specific case of `rotating drives' in $\Diff(S^1)$.
We provide an explicit formula for the resulting displacement between initial and final states, and use this to evaluate the return probability.
The consequence, in all cases, is a resonance pattern like the one in Fig.~\ref{introfig_loschmidt}.

%--------------------------------------------------------
\subsection{Return amplitudes from group theory}
\label{Sec:QGadiab:Amps}
%--------------------------------------------------------

We begin by showing how the Schr\"odinger equation with a time-dependent Hamiltonian \eqref{s11} can be solved by time-dependent coherent states of the form \eqref{e218}.
We then add the assumption of adiabaticity and periodicity to show that the return probability, \ie the squared complex norm of \eqref{s165}, probes the quantum metric \eqref{e226}.

\paragraph{From Schr\"odinger to paths in a group.}
Let $f_t$ be any family of deformations, and consider the time-dependent Hamiltonian $H(t) \equiv H[f_t]$ introduced in Eq.~\eqref{s11}.
We seek to solve the time-dependent Schr\"odinger equation 
\beq
i\partial_t|\psi(t)\rangle
=
H(t) |\psi(t)\rangle
=
\cU(f_t)H_0\cU(f_t)^{-1}|\psi(t)\rangle
\label{schro}
\eeq
with an initial state $|\psi(0)\rangle=\cU(f_0)|h\rangle$, chosen to be an eigenstate of the initial Hamiltonian.
[Recall from Eq.~\eqref{e217} that $|h\rangle$ is an eigenstate of $H_0$ with energy $\tfrac{2\pi}{L}(h-\tfrac{c}{24})$.]
The strategy is to reformulate the Schr\"odinger equation \eqref{schro} as an equation of motion in the group manifold by stripping off information about the Hilbert space and the representation $\cU$.
To this end, make the ansatz
\beq
|\psi(t)\rangle
=
e^{i\alpha_t} \cU(g_t)|h\rangle,
\label{ss17}
\eeq
where $\alpha_t \in \RR$ is a time-dependent overall phase (such that $\alpha_0 = 0$) and $g_t\in\Diff(S^1)$ is a path in the group manifold (such that $g_0=f_0$).
Both sides of Eq.~\eqref{schro} then take the form of operators acting on the highest-weight state $|h\rangle$.
Thus, Eq.~\eqref{schro} certainly holds if the operators themselves are equal, \ie if
\beq
-\der_t \alpha_t + i\bigl[\der_t\cU(g_t)\bigr] \cU(g_t)^{-1}
=
\cU(f_t)H_0\cU(f_t)^{-1}.
\label{s175}
\eeq
The right-hand side can be made more explicit thanks to the fact that $H_0$ itself is a Lie algebra generator.
Indeed, we already computed it in Eq.~\eqref{t11}.
For later convenience, we now write it as $\cU(f_t)H_0\cU(f_t)^{-1} = \frac{2\pi i}{L}\cu(\bV_t)+C[f_t]$, where
\beq
\bV_t
\equiv
V_t(\phii) \partial_{\phii}
\equiv
\frac{\der_\phii}{(f_t^{-1})^{\prime_{\vphantom{A}}}_{\vphantom{\tilde{A}}}(\phii)}
\label{ss175}
\eeq
is a time-dependent vector field.
Note that $V_t(\phii)$ is essentially the dimensionless velocity profile introduced in Eq.~\eqref{e11}: one has $V_t(\phii) = v_t(x)$ for $\phii = 2\pi x/L$.

This is where information on the operators $\cU(g)$ is needed.
As stated in Eq.~\eqref{UfUg_Vir}, they furnish a projective representation of $\Diff(S^1)$, so the left-hand side of Eq.~\eqref{s175} involves $[\partial_t\cU(g_t)] \cU(g_t)^{-1} = \cu(\der_tg_t \circ g_t^{-1}) + ic \partial_{s} \sfC(g_s, g_t^{-1})\big|_{s = t}$, where $\sfC(\cdot,\cdot)$ is the Bott cocycle \eqref{bott}.
We get rid of all terms proportional to the central charge $c$ by choosing the time-dependent phase $\alpha_t$ such that
\beq
-\partial_t \alpha_t - c\partial_{s} \sfC(g_s, g_t^{-1}) \big|_{s = t}
= C[f_t],
\label{t175}
\eeq
so what remains of Eq.~\eqref{s175} is $\cu(\der_tg_t \circ g_t^{-1}) = \tfrac{2\pi}{L}\cu(\bV_t)$.
This, in turn, is satisfied if
\beq
\label{evop}
\der_tg_t \circ g_t^{-1}
= \frac{2\pi}{L} V_t.
\eeq
The latter is an equation for the time-dependent diffeomorphism $g_t$ that no longer makes any reference to a Hilbert space or operators thereon.
If Eqs.~\eqref{evop} and \eqref{t175} hold, then so does the Schr\"odinger equation \eqref{schro} with the ansatz \eqref{ss17}.

A straightforward analogue of Eq.~\eqref{evop} holds in other situations involving generalized coherent states when the Hamiltonian preserves coherence [\ie when $H(t) = i\cu(\bV_t)$ for some Lie algebra element $\bV_t$].
The only difference lies in the choice of group manifold; see \eg \cite{PhysRevA.31.2721,Zalesny:2000zz} for the case of squeezed states.
For $G=\Diff(S^1)$ as in our current setup, Eq.~\eqref{evop} admits a hydrodynamical interpretation: $\bV_t$ can be seen as a time-dependent vector field that determines the local velocity of a (compressible) fluid on $S^1$, and the solution $g_t(\phii)$ of Eq.~\eqref{evop} gives the time-dependent position of a fluid parcel initially located at $g_0(\phii)=f_0(\phii)$.
Equivalently, $g_t$ is the one-parameter flow generated by $\bV_t$ (with respect to the dimensionless time $2\pi t/L$).

\begin{figure}[t]
\centering
\includegraphics[width=.4\textwidth]{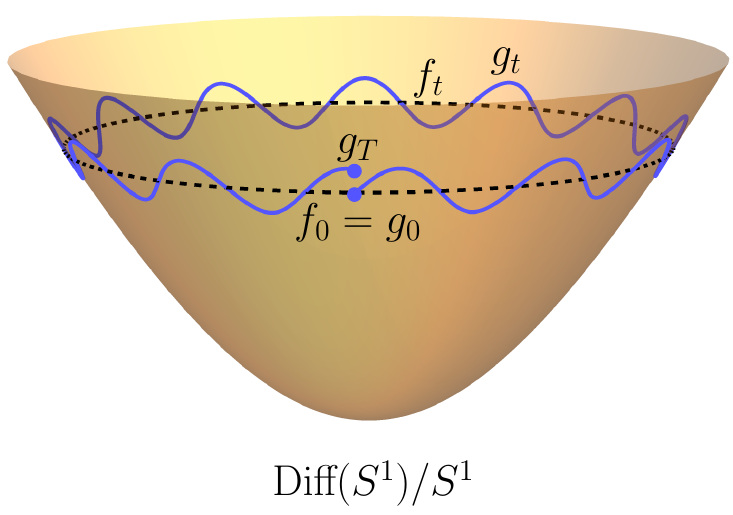}
\caption{%
Cartoon of the parameter space \eqref{s34}, with $G = \Diff(S^1)$ and $G_0 = S^1$ as in Eq.~\eqref{s125q}, showing the periodic drive $f_t$ and the corresponding solution $g_t$ of Eq.~\eqref{evop}.
In the adiabatic limit, the cosets $f_tG_0$ and $g_tG_0$ nearly coincide, but not quite, due to the small micromotion of $g_t$.
Their mismatch ultimately causes a nontrivial overlap $|\langle\psi(0)|\psi(T)\rangle| = |\langle h| \cU(g_0)^{-1} \cU(g_T) |h\rangle|\neq1$ that measures the quantum distance between nearby coherent states.
See also Fig.~\ref{Fig:Hyperbolicdisk} below.%
}%
\label{Fig:s41}
\end{figure}

\paragraph{Return probability and quantum metric.}
Let us now add two assumptions, namely (i) that the velocity vector field $\bV_t$ is periodic in time with period $T$, and (ii) that its time dependence is slow, or adiabatic, in the sense that the operator norm of $\der_t H(t)$ is much smaller than $1/L^2$.
In that limit, the adiabatic theorem states that the solution \eqref{ss17} sticks closely to $\cU(f_t) |h\rangle$, up to a time-dependent phase \cite{Born1928}.
Put differently, the coset of $g_t$ on the parameter space \eqref{s34} coincides with that of $f_t$.
The return amplitude \eqref{s165} is thus a pure phase to leading order, involving in particular a Berry phase.
As the latter is studied in detail in \cite{Oblak:2017}, we do not dwell on it here.
Our focus is instead on the first subleading correction to the adiabatic theorem, which accounts for a small difference between the cosets of $g_t$ and $f_t$ on the parameter space \eqref{s34}; see Fig.~\ref{Fig:s41}.
This leads to a nontrivial return probability given by the square of the amplitude \eqref{s165}.

To be specific, let $0<\epss\ll1$ denote a dimensionless adiabatic parameter, with $\epss \to 0$ indicating the adiabatic limit.
[The exact value of $\epss$ depends on the path $f_t$, but it is safe in general terms to think of it as $\epss \sim L/T$; see \eg Eq.~\eqref{omexp}.]
Let $G_0\cong S^1$ be the stabilizer that leaves the undeformed CFT Hamiltonian invariant, as in Eq.~\eqref{s125q}.
The small displacement between the cosets $g_t G_0$ and $f_t G_0$ in the parameter space \eqref{s34} can then be represented by a time-dependent Lie algebra element $\bY_t$ defined as follows: at each time $t$, there exists a finite rotation $R_{\theta(t)}(\phii) = \phii + \theta(t)$ by an angle $\theta(t)$ such that
\beq
f_t^{-1} \circ g_t
=
R_{\theta(t)} \circ e^{\epss\bY_t},
\label{s185}
\eeq
where $\bY_t$ is finite at all times in the adiabatic limit $\epss\to0$.
[Rotations appear here because they span the stabilizer of the undeformed Hamiltonian \eqref{e211}.]
Using this in the ansatz \eqref{ss17}, the return amplitude \eqref{s165} can be adiabatically expanded in $\epss \ll 1$ as
\begin{align}
& \langle\psi(0)|\psi(T)\rangle
=
e^{i\alpha_T + ic\sfC(f_0^{-1}\!,\,g_T)}\langle h|\cU(f_0^{-1}\circ g_T)|h\rangle \nonumber \\
& \sim
e^{i\Theta}
\biggl[ 1 - \frac{\epss^2}{2}\Bigl(\langle h|\bigl[i\cu(\bY_T)\bigr]^2|h\rangle - \langle h|i\cu(\bY_T)|h\rangle^2\Bigr) \biggr] \nonumber \\
& =
e^{i\Theta}
\biggl[ 1 - \frac{\epss^2}{2} \cG(\bY_T,\bY_T) \biggr],
\label{t185}
\end{align}
up to $O(\epss^3)$ corrections.
Here, $\Theta$ is an overall phase---the sum of a dynamical phase, a Berry phase \eqref{s43}, and subleading corrections in $L/T\sim\delta$.
For our purposes, the key part of Eq.~\eqref{t185} is the one that is \i{not} a phase, involving the squared quantum norm \eqref{e226} of $\bY_T$.\footnote{More precisely, the squared norm of $\bY_T + \mathfrak{g}_0$, seen as a tangent vector of the manifold \eqref{s34} at the identity.}
This gives rise to the return probability \cite{AnandanAharonov:1990}
\beq
\big|\langle\psi(0)|\psi(T)\rangle\big|^2
=
1 - \epss^2 \cG(\bY_T,\bY_T) + O(\epss^3)
\label{ss195}
\eeq
expressed in terms of the quantum metric, as announced in Eq.~\eqref{introlosch}.
Note that this is the metric at the identity, even though it measures the distance between the rays of $\cU(g_T)|h\rangle$ and $\cU(g_0)|h\rangle$, with $g_T$ and $g_0$ both well away from the identity.
This is again because the parameter space \eqref{s34} is homogeneous: its quantum geometry at the identity determines quantum geometry everywhere.

%--------------------------------------------------------
\subsection{Return probabilities for rotating drives}
\label{Sec:QGadiab:Probas}
%--------------------------------------------------------

We call \i{rotating drive} a family of time-dependent deformations given by
\beq
f_t(\phii) = f_0(\phii) + \omega t
\label{tt44}
\eeq
for some frequency $\omega\equiv2\pi/T>0$ and a fixed transformation $f_0 \in \Diff(S^1)$.
Any such transformations give rise to a periodic Hamiltonian \eqref{s11}.
Our goal is to find the resulting return probability \eqref{ss195} in the adiabatic limit $\omega L\to0$, where $L$ is system size.
We do this by first solving Eq.~\eqref{evop}, which turns out to be integrable in the case of rotating drives.
We then find the vector field $\bY_t$ in the group element \eqref{s185}, deduce the overlap $|\langle h|\cU(f_t^{-1}g_t)|h\rangle|^2$ at all times, and read off the return probability after one cycle.

\paragraph{Motion in parameter space for rotating drives.}
For a rotating drive \eqref{tt44}, the right-hand side of Eq.~\eqref{evop} is a vector field with profile $V_t(\phii) = V_0(\phii-\omega t)$, where $V_0(\phii) = 1/(f_0^{-1})'(\phii)$ owing to Eq.~\eqref{ss175}.
It is a local velocity that rotates around the circle at constant frequency $\omega$.
Let again $R_{\theta}(\phii)\equiv\phii+\theta$ denote rotation by $\theta$.
Then the flow $g_t$ corresponding to a rotating drive is readily found with the same ansatz as in \cite{Oblak:2020jek,Oblak:2020myi}.
Namely,
\beq
\label{gt}
g_t
=
R_{\omega t} \circ \zeta \circ R_{\Omega t} \circ \zeta^{-1} \circ g_0
\eeq
for a deformation $\zeta\in\Diff(S^1)$ and a frequency $\Omega$, both of which are time-independent unknowns.
The initial condition $g_0$ is arbitrary; choosing $g_0=f_0$ ensures that the initial state $|\psi(0)\rangle = \cU(f_0)|h\rangle$ is an eigenstate of the initial Hamiltonian, as in Sec.~\ref{Sec:QGadiab:Amps}.
Plugging the ansatz \eqref{gt} into Eq.~\eqref{evop} and using the winding condition \eqref{wind} yields
\beq
\label{zetaInv}
(\zeta^{-1})'(\phii)
=
\frac{\Omega}{\frac{2\pi}{L} V_0(\phii) - \omega},
\quad
\frac{1}{\Omega}
= \oint \frac{\dd\phii/2\pi}{\frac{2\pi}{L} V_0(\phii) - \omega},
\eeq
which entirely determines the flow \eqref{gt}.\footnote{Since $V_t(\phii)$ given by Eq.~\eqref{ss175} is continuous in $t$ and smooth in $\phii$, the existence and uniqueness theorem ensures that the ansatz \eqref{gt} is \i{the} solution of Eq.~\eqref{evop}.}
Note that this is well defined only if $V_0(\phii) - \frac{\omega L}{2\pi}$ never vanishes on the circle; in the adiabatic limit $\omega L\to 0$, this amounts to requiring that $V_0(\phii) \neq 0$,
which is indeed satisfied by any velocity profile of the form \eqref{ss175}.

Having found $g_t$, it is now straightforward to write the group element whose expectation value yields the return amplitude \eqref{t185}.
With the initial condition $g_0=f_0$, Eq.~\eqref{gt} yields
\beq
\label{finalrot}
f_0^{-1}\circ g_T
= f_0^{-1} \circ \zeta \circ R_{\Omega T} \circ \zeta^{-1} \circ f_0.
\eeq
This deformation is generally complicated---it is well away from the identity in $\Diff(S^1)$---but we are specifically interested in the adiabatic regime $\omega L \ll 1$ where things simplify.
To quickly see why, expand Eqs.~\eqref{zetaInv} in the adiabatic limit and keep only the leading terms, which yields $\Omega\sim2\pi/L$ and $(\zeta^{-1})'(\phii) \sim 1/V_0(\phii) = (f_0^{-1})'(\phii)$.
Thus, $\zeta \sim f_0$ in the adiabatic limit, so Eq.~\eqref{gt} implies $\lim_{\omega L \to 0} g_t = f_t \circ R_{2\pi t/L}$ and the group element \eqref{finalrot} boils down to a pure rotation $f_0^{-1} \circ g_T\sim R_{2\pi T/L}$.
The latter gives rise to the expected dynamical phase in the amplitude \eqref{t185}.
Our goal is to go beyond leading order by finding the small deviation of Eq.~\eqref{finalrot} away from a rotation.

\paragraph{Adiabatic expansion.}
Since $\Omega\sim2\pi/L$ in the adiabatic limit, let us set up the adiabatic expansion by defining the `effective' adiabatic parameter
\beq
\label{epss_exact}
\epss
\equiv
\frac{\omega}{\Omega}
=
\frac{1}{2\pi} \oint \frac{(L/T) \, \dd\phii}{V_0(\phii)-L/T}
\ll 1,
\eeq
where $\Omega$ was introduced in Eq.~\eqref{zetaInv}.
This can be contrasted with the `bare' adiabatic parameter $\omega L/2\pi = L/T$.
The two are related by the adiabatic expansion of Eq.~\eqref{zetaInv}, namely
\begin{align}
& \frac{1}{\epss}
=
\frac{T}{L}
- \oint \frac{\dd\phii}{2\pi f_0'(\phii)} \label{omexp} \\
& \;\;\; - \frac{L}{T} \bigg[\oint\frac{\dd\phii}{2\pi f_0'(\phii)^2}-\left(\oint\frac{\dd\phii}{2\pi f_0'(\phii)}\right)^2\,\bigg]
+ O\Bigl(\frac{L^2}{T^2}\Bigr), \nonumber
\end{align}
confirming that $\epss\sim L/T$ to leading order.
[We used $V_0(\phii) = 1/(f_0^{-1})'(\phii)$ from Eq.~\eqref{ss175}.]
In practice, the first two terms of the expansion \eqref{omexp} often suffice.
They are reminiscent of a dynamical phase and a Berry phase, to which they will be seen to correspond for $\mathrm{SL}(2,\RR)$ drives in Sec.~\ref{Sec:SL2R}.

It remains to compute the deformation $\zeta^{-1}\circ f_0$ of Eq.~\eqref{finalrot}.
The latter is close to the identity in the adiabatic limit: the expansion of Eq.~\eqref{zetaInv} in powers of $L/T$ yields
\begin{gather}
\label{hhf}
\zeta^{-1}(f_0(\phii))
=
\phii + \epss W(\phii) + O(\epss^2), \\
W(\phii)
\equiv
\int_0^{\phii} \dd\alpha 
\left(\frac{1}{f_0'(\alpha)}-\oint\frac{\dd\theta}{2\pi f_0'(\theta)}\right), \nonumber
\end{gather}
where we used Eq.~\eqref{omexp} to trade $L/T$ for $\epss$.
As expected, this is an infinitesimal diffeomorphism involving a vector field $\epss W(\phii)\der_\phii$.\footnote{This $\bW$ has nothing to do with the one used for linear response in Sec.~\ref{Sec:PertQG:LinResp}.}
One can now plug $\zeta^{-1}\circ f_0$ into Eq.~\eqref{gt} with $g_0=f_0$ to find
\beq
\label{Fpert}
\begin{aligned}
f_t^{-1}(g_t(\phii))
& = \phii + \Omega t + \epss Y_t(\phii) + O(\epss^2), \\
Y_t(\phii)
& \equiv W(\phii) - W(\phii+\Omega t).
\end{aligned}
\eeq
This is indeed a diffeomorphism of the form identified in Eq.~\eqref{s185}, with a rotation by $\theta(t)=\Omega t$ and a small $O(\epss)$ correction involving the time-dependent vector field $\bY_t = Y_t(\phii) \partial_{\phii}$.
The latter can be expressed using $f_0(\phii)$ thanks to Eq.~\eqref{hhf}.
Note that this holds at all times; at the end of one cycle, one has $\Omega t=\Omega T=2\pi/\epss$ in terms of the adiabatic parameter \eqref{epss_exact}.

\paragraph{Return probabilities.}
The probability
\eqref{ss195} is entirely determined by the Lie algebra element $\bY_T$, which we just found in Eqs.~\eqref{hhf}--\eqref{Fpert}.
Using the quantum metric written in Eq.~\eqref{cG_XY}, we conclude that
\begin{align}
& \big|\langle\psi(0)|\psi(T)\rangle\big|^2
\sim
1
- \epss^2 \oint\frac{\dd\phii\,\dd\theta}{16\pi^2 \sin^2([\phii-\theta]/2)} \nonumber \\
& \times \Biggl[
  \Bigl( h-\frac{c}{24} \Bigr) \biggl(
    \int_{\theta}^{\theta+\frac{2\pi}{\epss}}\!\!\!\frac{\dd\alpha}{f_0'(\alpha)} - \int_{\phii}^{\phii+\frac{2\pi}{\epss}}\!\!\!\frac{\dd\alpha}{f_0'(\alpha)}
  \biggr)^2 \nonumber \\
& \;\;\; +
    \frac{c}{24} \biggl(
      \frac{f_0'(\theta){-}f_0'(\theta+\frac{2\pi}{\epss})}{f_0'(\theta)f_0'(\theta+\frac{2\pi}{\epss})}
      - \frac{f_0'(\phii){-}f_0'(\phii+\frac{2\pi}{\epss})}{f_0'(\phii)f_0'(\phii+\frac{2\pi}{\epss})}
    \biggr)^2
\Biggr]
\label{cL_Vir_final}
\end{align}
for adiabatic rotating drives in CFT, where we used $\Omega T=2\pi/\epss$.
This is plotted in Fig.~\ref{introfig_loschmidt} as a function of $T/L \sim 1/\delta$ for different initial deformations $f_0$.

A striking aspect of the plots in Fig.~\ref{introfig_loschmidt} is the oscillating convergence of the return probability in the adiabatic limit $T/L\to\infty$.
These oscillations stem from the ubiquitous appearance, in Eq.~\eqref{cL_Vir_final}, of the shift by $2\pi/\epss\sim T/L$ in the $2\pi$-periodic function $1/f_0'(\phii)$: depending on the value of $2\pi/\epss$, the shift gives rise to constructive or destructive `interferences' of $1/f_0'(\phii)$.
A more direct way to predict such oscillations is to refer to the expression \eqref{cG_XmYn} of the Virasoro quantum metric in terms of Fourier modes.
In the case at hand, the Fourier modes of $Y_t(\phii)$ in Eq.~\eqref{hhf} are $\hat{Y}_m(t)=\hat{W}_m(1-e^{im\Omega t})$, where $\hat{W}_m$ is the $m^{\text{th}}$ mode of $1/f_0'(\phii)$ (for $m\neq0$).
The exponential $e^{im\Omega t}$ leads to an oscillating pattern of the squared quantum norm 
\begin{multline}
\label{metryc}
\cG(\bY_t,\bY_t)
=
4\sum_{m\in\ZZ} |m|\Big(h + \frac{c}{24} (m^2-1)\Big) \\ 
\times |\hat{W}_{m}|^2\sin^2(m\Omega t/2)
\end{multline}
as a function of $t$, reflecting the micromotion of $g_t$ around $f_t$ seen in Fig.~\ref{Fig:s41}.
This micromotion is manifest in the time-dependent overlap $|\langle h| \cU(f_t^{-1}\circ g_t) |h\rangle |^2 \sim 1 - \delta^2 \cG(\bY_t,\bY_t)$ between the exact state $\propto \cU(g_t)|h\rangle$ and its adiabatic approximation $\propto \cU(f_t \circ R_{2\pi t/L})|h\rangle$.

\paragraph{Adding left movers.}
The chiral case is readily generalized to a nonchiral CFT that involves both right and left movers, as discussed around Eq.~\eqref{cot} and in Appendix~\ref{App:CFTdict}.
This is crucial for the effective description of lattice models and for the comparison with their numerics, such as those in Fig.~\ref{introfig_loschmidt}.

The Hamiltonian of a driven nonchiral inhomogeneous CFT is a sum
\beq
\label{fullh}
H_{\text{nonchiral}}(t)
=
\int \dd x\,v_t(x)\big[\stt(x)+\overline\stt(x)\big]
\eeq
of the right-moving term \eqref{e11} with its left-moving partner.
The same velocity profile $v_t(x)$ multiplies both light-cone components of the stress-energy tensor, namely $\stt(x) = (2\pi/L)^2 \stt(\phii)$ and $\overline{\stt}(x) = (2\pi/L)^2 \overline{\stt}(\phii)$ for $\phii = 2\pi x/L$.\footnote{One could use different velocity profiles for right and left movers. This is more complicated to implement in lattice models, so we do not discuss such generalizations.}
While the Fourier expansion of $\stt$ is provided by Eq.~\eqref{e28} in terms of right-moving Virasoro generators $L_m$, that of $\overline{\stt}$ is
\beq
\label{e28bar}
\overline\stt(\phii)
=
\frac{1}{2\pi}
\sum_{n\in\ZZ} e^{-in \phii}
\left( \bar{L}_n - \frac{\bar{c}}{24} \delta_{n,0} \right)
\eeq
in terms of left-moving Virasoro generators $\bar{L}_m$.
The latter commute with the $L_m$s and satisfy the same commutation relations \eqref{e24} of the Virasoro algebra, but with a left-moving central charge $\bar{c}$ that may differ from $c$.
The nonchiral Hilbert space is a sum of tensor products of right and left representations of the Virasoro algebra, respectively generated by the $L_m$ and $\bar{L}_m$ operators.
Nonchiral highest-weight states are tensor products of the chiral ones in Eq.~\eqref{e217}.

The fact that right and left sectors commute means that the full return amplitude \eqref{s165} in a nonchiral CFT is the product of right- and left-moving contributions that can be computed separately.
The same is true of the return probability \eqref{ss195}.
In the same adiabatic expansion as before, the latter involves again the quantum metric, which is now a sum of contributions from right and left movers:
\beq
\big|\langle\psi(0)|\psi(T)\rangle\big|^2
\sim
1 - \epss^2 \cG(\bY_T,\bY_T) - \bar{\epss}^2 \bar{\cG}(\bar{\bY}_T,\bar{\bY}_T).
\label{ss195bis}
\eeq
Here all barred quantities denote left-moving analogues of (and generally differ from) the right-moving ones, and the state $|\psi(t)\rangle$ is that of the full nonchiral theory.
In particular, $\bar{\cG}$ has formulas analogous to Eqs.~\eqref{cG_XmYn}--\eqref{cG_mn} or \eqref{cG_XY} with $c$ and $h$ replaced by $\bar{c}$ and $\bar{h}$, where $\bar{h}$ is the left-moving highest weight.
The fact that the same velocity profile multiplies both $\stt$ and $\overline\stt$ in the driven Hamiltonian \eqref{fullh} further means that the left-moving flow $\bar g_t(\phii)$ 
satisfies an almost identical equation to Eq.~\eqref{evop} for the right-moving flow $g_t(\phii)$, the only difference being a sign:
\beq
\der_t\bar g_t \circ \bar g_t^{-1}
= - \frac{2\pi}{L} V_t.
\label{flowbis}
\eeq
The sign difference stems from the opposite directions of the flows, echoed by sign differences in many other formulas between right and left movers (see Appendix~\ref{App:CFTdict}).
This includes the different conventions for the Fourier expansions \eqref{e28} and \eqref{e28bar}, ensuring that the $L_m$s and $\bar{L}_m$s still satisfy the same commutations relations \eqref{e24}.

For rotating drives where $V_t(\phii)=V_0(\phii-\omega t)$, the fact that left movers experience a velocity \textit{opposite} to that of right movers has important implications for the return probability \eqref{ss195bis}.
An analogue of Eq.~\eqref{gt} still holds, namely $\bar g_t=R_{\omega t}\circ \bar\zeta\circ R_{-\overline\Omega t}\circ \bar\zeta^{-1}\circ\bar g_0$, but solving the left-moving flow equation \eqref{flowbis} now yields
\beq
(\bar\zeta^{-1})'(\phii)
=
\frac{\overline\Omega}{\frac{2\pi}{L} V_0(\phii) + \omega},
\quad
\frac{1}{\overline\Omega}
= \oint \frac{\dd\phii/2\pi}{\frac{2\pi}{L} V_0(\phii) + \omega}.
\label{bb21b}
\eeq
These only differ from the right-moving formulas in Eq.~\eqref{zetaInv} by the sign in front of $\omega$.
In particular, the left-moving adiabatic parameter is $\bar\epss \equiv \omega/\overline\Omega$, which differs from Eq.~\eqref{epss_exact} by a sign in its denominator,
and the Lie algebra element $\bar{\bY}_T$ for rotating drives can be found just as we did for $\bY_T$ in Eqs.~\eqref{hhf}--\eqref{Fpert}.
The return probability \eqref{ss195bis} is given by the product of Eq.~\eqref{cL_Vir_final} with its left-moving cousin, obtained by replacing $\epss$, $c$, $h$ by $\bar{\epss}$, $\bar{c}$, $\bar{h}$.
Equivalently, it can be Fourier expanded in a way analogous to Eq.~\eqref{metryc}:
\begin{align}
\label{sss}
& \big|\langle\psi(0)|\psi(T)\rangle\big|^2
\sim
1 \\
& - 4\epss^2\sum_{m\in\ZZ} |m|
    \Bigr(h + \frac{c}{24} (m^2{-}1) \Bigl) |\hat{W}_{m}|^2
    \sin^2 \Bigl(\frac{m\Omega T}{2}\Bigr) \nonumber \\
& - 4\bar\epss^2\sum_{m\in\ZZ} |m|
    \Bigl(\bar{h} + \frac{\bar{c}}{24} (m^2{-}1)\Bigr) |\hat{W}_{m}|^2
    \sin^2 \Bigl(\frac{m\overline{\Omega} T}{2}\Bigr). \nonumber
\end{align}
The CFT used in Fig.~\ref{introfig_loschmidt} has $c = \bar{c} = 1$, and the initial state $|\psi(0)\rangle$ is the (deformed) nonchiral ground state with $h = \bar{h} = 0$.

Note an important difference between the nonchiral formula \eqref{sss} and its right-moving simplification \eqref{metryc}.
The latter vanishes at $t=T$ when $\Omega T$ is an integer multiple of $2\pi$, so the chiral return probability \eqref{cL_Vir_final} is guaranteed to reach the value $|\langle\psi(0)|\psi(T)\rangle|^2 = 1$ repeatedly when $T$
tends to infinity.
By contrast, the nonchiral probability \eqref{sss} never quite reaches that value because $\Omega\neq\overline\Omega$, even though $\Omega \sim 2\pi/L \sim \overline{\Omega}$ in the adiabatic limit.
The consequence of this slight mismatch between $\Omega$ and $\overline{\Omega}$ is visible \eg in Fig.~\ref{introfig_loschmidt}(b), where the maxima of the return probability all sit slightly yet strictly below $1$.

%==========================================================
\section{Rotating drives in \texorpdfstring{$\mathrm{\bf SL}\boldsymbol(\boldsymbol2\boldsymbol,\mathbb{R}\boldsymbol)$}{SL(2,R)}}
\label{Sec:SL2R}
%==========================================================

The simplest case of an inhomogeneous Hamiltonian \eqref{e11} has a velocity profile involving a single wavenumber $k$, so that $v_t(x)=A_t+B_t\cos(2\pi kx/L-C_t)$ with generally time-dependent coefficients $A_t > |B_t|$, $B_t$, and $C_t$.
In contrast to the protocols of Secs.~\ref{Sec:PertQG} and~\ref{Sec:QGadiab}, such drives do not probe the full parameter space of conformal frames.
They are limited, instead, to the subgroup of $\Diff(S^1)$ that consists of transformations of the form
\beq
e^{ik f(\phii)}
=
\frac{\alpha e^{ik \phii} + \beta}{\bar\beta e^{ik \phii} + \bar\alpha}
\label{s435}
\eeq
for $\alpha, \beta \in \CC$ satisfying $|\alpha|^2 - |\beta|^2 = 1$.
This subgroup is finite-dimensional and locally isomorphic to $\mathrm{SL}(2,\RR) \cong \mathrm{SU}(1,1)$.
As a result, the dynamics of coherent states can be wholly expressed in terms of finite-dimensional matrices.
We now study one such situation with a rotating drive $v_t(x) = A + B\cos(2\pi kx/L-\omega t)$ (where $A > |B|$ and $B$ are constants) and show that the evolution operator can be computed \i{exactly} for any driving frequency $\omega>0$ and general amplitudes $A$, $B$.
This, in turn, allows us to compute nonlinear responses and return amplitudes in any driving regime, as illustrated in Fig.~\ref{fiReg}.

%--------------------------------------------------------
\subsection{Evolution operator}
\label{Sec:SL2R:Evol}
%--------------------------------------------------------

Pick an integer wavenumber $k$ and a real `rapidity' parameter $\lambda>0$.
Given the component $\stt(\phii)$ of the CFT stress-energy tensor, consider the time-dependent inhomogeneous Hamiltonian
\begin{widetext}
\beq
\label{Ht_SL2}
H(t)
=
\frac{2\pi}{L} \oint \dd\phii\, \Bigl[ \cosh(2\lambda) - \sinh(2\lambda) \cos(k[\phii - \omega t]) \Bigr] \stt(\phii)
+ \frac{\pi c}{12 L} \bigl[ \cosh(2\lambda) - 1 \bigr].
\eeq
This takes the form \eqref{t11} for a rotating drive \eqref{tt44} whose `initial deformation' $f_0$ is given by Eq.~\eqref{s435} with $\alpha = \cosh(\lambda)$ and $\beta = \sinh(\lambda)$.
Equivalently,
\beq
H(t)
=
\frac{2\pi}{L} \biggl[ \cosh(2\lambda)L_{0} - \frac{1}{2} \sinh(2\lambda) \Bigl( e^{ik\omega t} L_{k} + e^{-ik\omega t} L_{-k} \Bigr) \biggr]
+ \frac{\pi c}{12 L} \bigl[ (k^2 - 1)\cosh(2\lambda) - 1 \bigr]
\label{eq:Ht_SL2_Ls}
\eeq
\end{widetext}
using the Virasoro generators $L_{n}$ in Eqs.~\eqref{e28}--\eqref{e24}.
The Hamiltonian thus involves only $\{ L_{-k}, L_{0}, L_{k} \}$, which generate an $\mathfrak{sl}(2,\RR) \cong \mathfrak{su}(1,1)$ subalgebra (see Appendix~\ref{App:SL2R}).
It follows that the evolution operator
\beq
\label{Ut_def}
U(t)
\equiv
\overset{\longleftarrow}{\cT} \exp \left( -i \int_0^t \dd t'\, H(t') \right),
\eeq
with ordering $\overset{\longleftarrow}{\cT}$ so that time increases from right to left, belongs to the corresponding $\mathrm{SL}(2,\RR) \cong \mathrm{SU}(1,1)$ subgroup of $\Diff(S^1)$.
Since this is a matrix group, one may hope to use matrix multiplication to compute $U(t)$, despite the nontrivial time ordering.
No such simplification would be available in the full Virasoro algebra.

To obtain the evolution operator \eqref{Ut_def}, write the time-ordered exponential as
\beq
\label{Ut_Trotter}
U(t)
=
\lim_{M\to\infty} \prod_{m=0,\ldots,M}^{\longleftarrow} e^{-iH(m\delta t)\delta t},
\eeq
where $\delta t\equiv t/M$ and the product $\overset{\longleftarrow}{\prod}$ is ordered so that time increases from right to left.
Each factor in this product can be viewed as (a unitary representation of) a matrix in $\mathrm{SU}(1,1)$.
Thus, one may just as well deal with the matrices directly, and evaluate their product.
This is explained in Appendix~\ref{App:SL2R}, to which we refer for details.
Suffice it to say that a key simplification occurs thanks to the drive being a rotating one, in the sense of Eq.~\eqref{tt44}: the $m$ dependence of each factor in \eqref{Ut_Trotter} cancels out against the $(m+1)$ of the next factor, and the entire evolution operator \eqref{Ut_Trotter} becomes the $M^{\text{th}}$ power of a time-independent matrix.
The end result is
\begin{align}
\label{Ut_SL2_Ls}
& U(t)
=
e^{-i \frac{\pi c}{12 L} k^2 t [\cosh(2\lambda) - 1]}
e^{-i \omega t (L_0 - c/24)} \\
& \quad \times
e^{-i \frac{2\pi}{L} t \bigl( [ \cosh(2\lambda) - L/T] (L_{0} - c/24) - \sinh(2\lambda) \frac{L_{-k} + L_{k}}{2} \bigr)}, \nonumber
\end{align}
expressed in terms of the Virasoro generators.
Note the structure of this expression, which is consistent with the evolution operator for coherent states in Eq.~\eqref{ss17} and the specific form \eqref{gt} of $g_t$ for rotating drives.
Indeed, using $e^{i\alpha_t} \cU(g_t) = U(T) \cU(g_0)$, one can identify the second exponential in Eq.~\eqref{Ut_SL2_Ls} with the undeformed rotation $R_{\omega t}$, while the last exponential in Eq.~\eqref{Ut_SL2_Ls} can be shown to represent the `deformed rotation' $\zeta \circ R_{\Omega t} \circ \zeta^{-1}$.
In the case at hand, $\zeta$ is an $\mathrm{SU}(1,1)$ deformation of the form \eqref{s435} with (see again Appendix~\ref{App:SL2R})
\beq
\alpha = \frac{1+\Lambda}{2\sqrt{\Lambda}},
\quad\;
\beta = \frac{1-\Lambda}{2\sqrt{\Lambda}},
\quad\;
\Lambda \equiv 2\pi \frac{e^{-2\lambda} - L/T}{L\Omega},
\label{alphabeta_SL2_zeta}
\eeq
and the frequency of the rotation is
\beq
\label{Omega_SL2}
\Omega
=
\frac{2\pi}{L} \sqrt{ 1 - 2\cosh(2\lambda) (L/T) + (L/T)^2 },
\eeq
all of which is consistent with Eq.~\eqref{zetaInv}.
What is special about $\mathrm{SL}(2,\RR)$ drives is that the deformed rotation can be written as an explicit exponential of finitely many Virasoro generators.
This would not be true with more general rotating drives, as we saw in Sec.~\ref{Sec:QGadiab}.

\begin{figure*}[!ht]
\centering
\includegraphics[width=.6\textwidth]{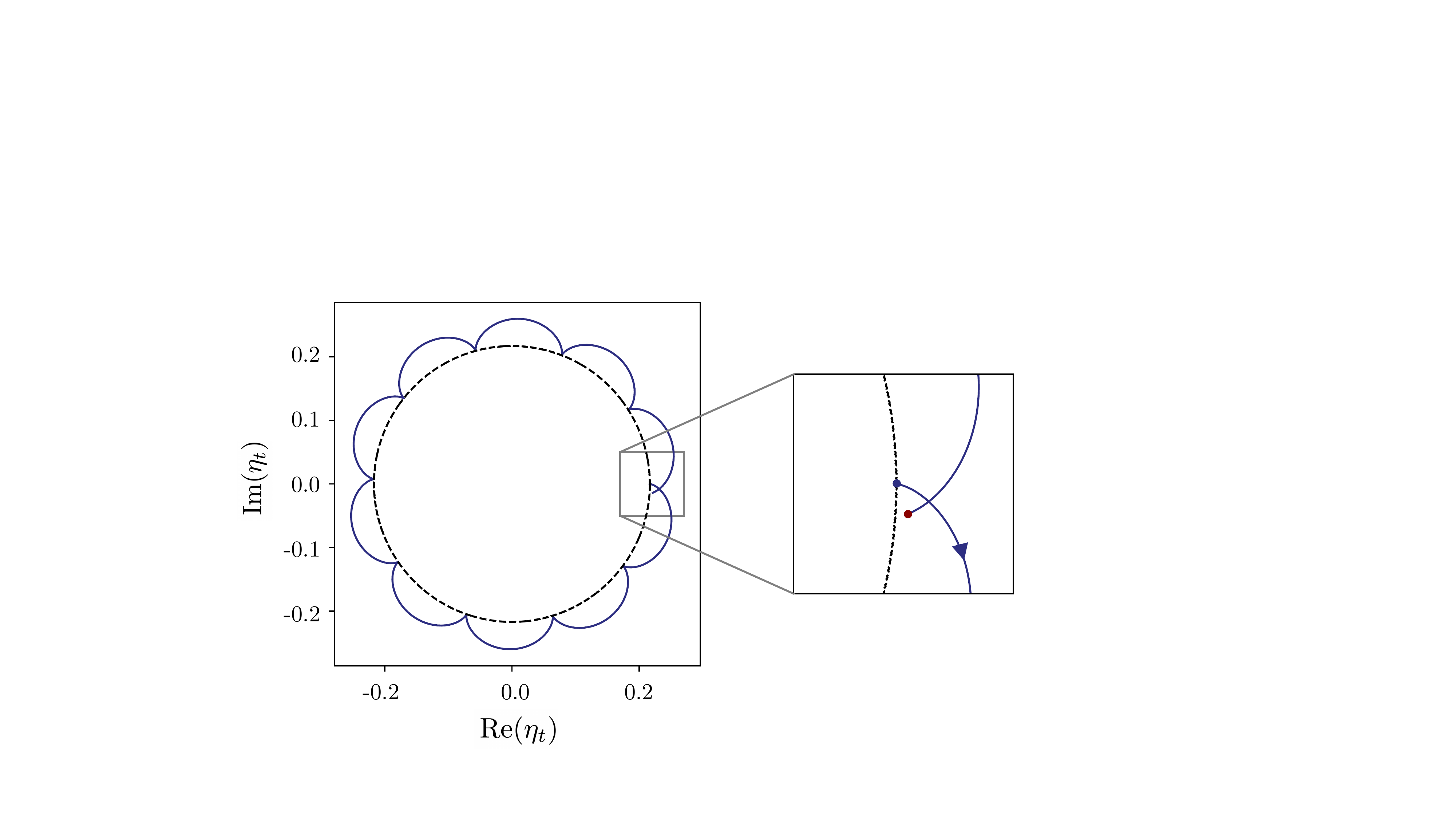}
\caption{%
Time evolution of the unit disk coordinate $\eta_t$ in Eq.~\eqref{eta_t_SL2} induced by the dynamics given by the Hamiltonian \eqref{eq:Ht_SL2_Ls} with $k = 1$ and $\lambda = 0.22$.
The result $\tanh(\lambda) e^{-ik\omega t}$ in the adiabatic limit $L/T \to 0$ for the trajectory $\eta_t$ is shown as a black dashed line, while the blue curve shows the nonadiabatic result for $L/T = 1/11$.
After one full period, the final state (red dot) does not exactly return to the initial state (blue dot).
The slight mismatch between the corresponding coherent states gives rise to a nearly saturated return probability.%
}%
\label{Fig:Hyperbolicdisk}
\end{figure*}

\paragraph{Motion on parameter space.}
Since the solution \eqref{ss17} of the Schr\"odinger equation consists of coherent states, the dynamics encoded by the evolution operator \eqref{Ut_SL2_Ls} defines a path in the parameter space \eqref{s34}: at time $t$, the state vector $|\psi(t)\rangle=e^{i\alpha_t} \cU(g_t)|h\rangle$ defines a unique coset $g_tG_0\in G/G_0$.
(Conversely, the coset $g_tG_0$ uniquely determines the ray $[|\psi(t)] = [\cU(g_t)|h\rangle]$.)
The situation is especially simple for $\mathrm{SL}(2,\RR)$ drives, which only involve three Virasoro generators $\{ L_{-k}, L_{0}, L_{k} \}$ and therefore probe a two-dimensional parameter space $\mathrm{SL}(2,\RR)/S^1$ (\ie a hyperbolic plane).
One can thus view the motion of the ray $[|\psi(t)\rangle]$ as a curve on a hyperbolic disk, as in Fig.~\ref{Fig:Hyperbolicdisk}.
Concretely, the curve is obtained by writing the ray of the state vector as
\beq
\label{psi_t_factorized}
\bigl[|\psi(t)\rangle\bigr]
=
\bigl[e^{ \chi_t L_{-k} - \bar{\chi}_t L_{k} } |h\rangle\bigr]
\eeq
for $\chi_t$ a time-dependent complex function that can be conveniently parametrized as $\chi_t \equiv \frac{1}{k}\operatorname{artanh}(|\eta_t|)\frac{\eta_t}{|\eta_t|}$ in terms of $\eta_t$ on the unit (Poincar\'e) disk.
One finds (see Appendix~\ref{App:SL2R})
\begin{align}
\label{eta_t_SL2}
\eta_t
& \equiv
\tanh(k|\chi_t|) \frac{\chi_t}{|\chi_t|} \\
& =
\tanh(\lambda)\,
\frac{
  1 + i \Bigl( \frac{2\pi}{L\Omega} + \frac{\omega}{\Omega} \Bigr) \tan(k\Omega t/2)
}{
  1 + i \Bigl( \frac{2\pi}{L\Omega} - \frac{\omega}{\Omega} \Bigr) \tan(k\Omega t/2)
}
\, e^{-i k\omega t} \nonumber
\end{align}
with $\Omega$ the frequency \eqref{Omega_SL2}.
This is illustrated in Fig.~\ref{Fig:Hyperbolicdisk}.
Note the behavior in the adiabatic limit, where $\lim_{L/T \to 0} \eta_t = \tanh(\lambda) e^{-i k\omega t}$ essentially coincides with the rotating drive, as was to be expected.
In fact, the complex variable $\eta_t$ is known to satisfy a Riccati equation with time-dependent coefficients determined by the external drive \cite{PhysRevA.31.2721}, but this perspective is not needed in our case.

%--------------------------------------------------------
\subsection{Nonlinear and linear responses}
\label{Sec:SL2R:Resps}
%--------------------------------------------------------

A first application of the evolution operator \eqref{Ut_SL2_Ls} is to compute the effect of the driven Hamiltonian \eqref{Ht_SL2} on any operator \eqref{e27} generated by the stress-energy tensor, similarly to what we did in Sec.~\ref{Sec:PertQG} to linear order.
The result is actually quite general: provided $g_t$ is chosen such that the state vector $|\phi(t)\rangle = e^{i\alpha_t} \cU(g_t)|h\rangle$ solves the Schr\"odinger equation \eqref{schro} with initial condition $|\phi(0)\rangle = \cU(g_0)|h\rangle$, the corresponding expectation value of $i\cu(\bW)$ is
\begin{align}
\label{nonlin_resp_gen}
& \langle \phi(t)| i\cu(\bW) |\phi(t)\rangle
=
\frac{1}{2\pi} \oint \dd\phii\, W(\phii) \\
& \quad \times
\biggl(
\Bigl( h - \frac{c}{24} \Bigr)
\bigl[ (g_t^{-1})'(\phii) \bigr]^2
- \frac{c}{12}\{ g_t^{-1}, \phii \}
\biggr). \nonumber
\end{align}
This follows from the transformation law \eqref{tranT} of the CFT stress tensor, and it holds regardless of the initial $g_0$.
In the case of rotating drives, Eq.~\eqref{gt} allows one to express $\langle \phi(t)| i\cu(\bW) |\phi(t)\rangle$ in terms of the deformation $\zeta(\phii)$, the frequency $\Omega$, and $g_0$.

For the specific case of rotating drives in $\mathrm{SL}(2,\RR)$, this gives explicit access to the full nonlinear response.
Indeed, $\zeta(\phii)$ is known to be of the form \eqref{s435} with $\alpha$, $\beta$ in Eq.~\eqref{alphabeta_SL2_zeta}, and $\Omega$ is also known exactly from Eq.~\eqref{Omega_SL2}.
Choosing $g_0$ to be the identity, and considering the special case of small perturbations $\epsilon = \lambda \ll 1$, it follows that
\begin{align}
& \langle \phi(t)| i\cu(\bW) |\phi(t)\rangle 
=
\Bigl( h - \frac{c}{24} \Bigr) \hat{W}_0 \\
& \quad \begin{aligned}
& - i \epsilon \cF_{k,-k} \biggl( \hat{W}_{k} \frac{e^{ik\omega t} - e^{ik2\pi t/L}}{k(1 - L/T)} \\
& \qquad\qquad\quad + \hat{W}_{-k} \frac{e^{-ik\omega t} - e^{-ik2\pi t/L}}{k(1 - L/T)} \biggr)
+ O(\epsilon^2)
\end{aligned} \nonumber
\end{align}
in terms of the Berry curvature $\cF_{m,n}$ of Eq.~\eqref{cF_mn}.
This agrees with the general linear-response result \eqref{htcOht_suscep}--\eqref{chi_mn} by noting that the $\bX_t$ in Eq.~\eqref{Ht_uXt} is now given by $X_t(\phii) = - \frac{2}{k} \sin(k[\phii - \omega t])$.
In fact, the general expression \eqref{nonlin_resp_gen} is consistent with our linear-response formulas \eqref{htcOht_suscep}--\eqref{chi_mn}: this can be shown by expressing $g_t(\phii)$ in terms of a weakly perturbed $f_t(\phii) = \phii + \epsilon X_t(\phii)$.

%--------------------------------------------------------
\subsection{Return amplitudes and probabilities}
\label{Sec:SL2R:AmpsProbas}
%--------------------------------------------------------

A second application of Eq.~\eqref{Ut_SL2_Ls} is to find the solution $U(t)|\psi(0)\rangle$ of the Schr\"odinger equation \eqref{schro} for any initial condition $|\psi(0)\rangle$.
As in Sec.~\ref{Sec:QGadiab}, we choose $|\psi(0)\rangle=\cU(f_0)|h\rangle$ to be an eigenstate of the initial inhomogeneous Hamiltonian, and investigate the return amplitude \eqref{s165}.
This can again be done in terms of $2\times2$ matrices, and the return amplitude can in fact be evaluated for any time $t$ (as opposed to $t=T$ at the end of one cycle).
In terms of the convenient parameter $h_k\equiv h+\tfrac{c}{24}(k^2-1)$, one finds (see Appendix~\ref{App:SL2R})
\beq
\label{p0pt_SL2}
\langle \psi(0)|\psi(t) \rangle
=
e^{i\pi c k^2t/12L}
u_t^{-2h_{k}/k}
\eeq
with
\begin{align}
\label{eq:u_t}
u_t
& = \cos(k\Omega t/2) \cos(k\omega t/2) \\
& \quad - 2\pi \frac{\cosh(2\lambda) - L/T}{L\Omega} \sin(k\Omega t/2)\sin(k\omega t/2) \nonumber \\
& \quad + i \cosh(2\lambda) \biggl[ \cos(k\Omega t/2)\sin(k\omega t/2) \nonumber \\
& \qquad + 2\pi \frac{\operatorname{sech}(2\lambda) - L/T}{L\Omega} \sin(k\Omega t/2)\cos(k\omega t/2) \biggr] \nonumber 
\end{align}
expressed in terms of the frequency \eqref{Omega_SL2}.
The overall phase of $\langle \psi(0)|\psi(t) \rangle$ can be extracted by picking the branch of $\arg(u_t)$ consistent with the limiting case $\lambda = 0$, for which the phase is purely dynamical.
This generalizes previous results for the return probability in $\mathfrak{sl}(2,\RR)$-driven CFT \cite{PhysRevResearch.2.033461}, which were restricted to piecewise-constant time-dependent Hamiltonians.
Furthermore, the return probability was recently studied for continuous Floquet drives in \cite{Das_2021_floquetCFT}, which relied on Floquet perturbation theory, while the result \eqref{p0pt_SL2} is exact and uses no perturbation theory.

\begin{figure*}[!ht]
\centering
\includegraphics[width=.9\textwidth]{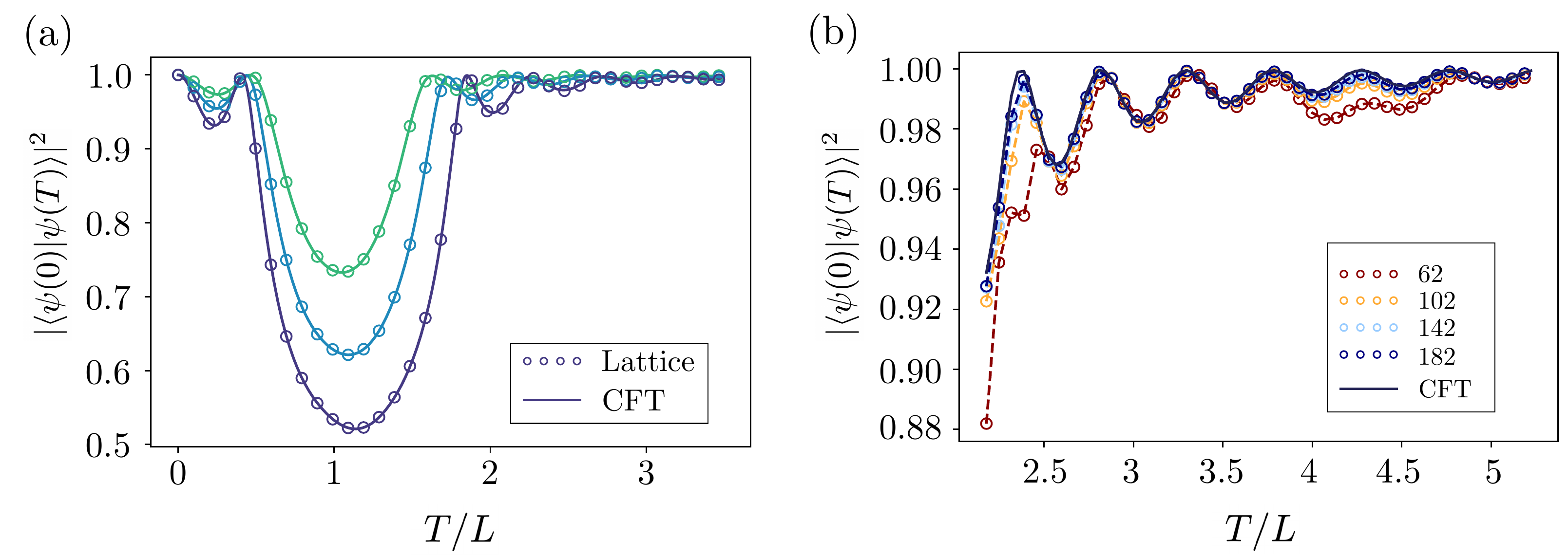}
\caption{%
(a) Return probability $|\langle \psi(0)|\psi(T) \rangle|^2$ after one period $T$
of the Hamiltonian \eqref{eq:Ht_SL2_Ls}, starting from the ground state $|\psi(0)\rangle$.
The results are plotted as a function of $T/L$ for $\lambda = 0.15,0.2,0.25$ (green to purple).
(The parameters are $L = 202$, $k=2$, $c = 1$, $h = 0$ and periodic boundary conditions are imposed.)
We compare the exact CFT result with numerical lattice calculations for a free-fermion chain at half filling with $N = L$ sites (unit lattice spacing) and inhomogeneous hoppings of the form \eqref{Jj_t_SL2}
(see Appendix~\ref{App:LattCalcs}).
The CFT result is the product of the squared complex norm of Eq.~\eqref{p0pt_SL2} with its left-moving counterpart; the latter is obtained from the same formula with $\omega$ replaced by $-\omega$ everywhere---including in $\Omega$, which is thus replaced by $\overline\Omega$ in Eq.~\eqref{ttt25b}.
(b) Finite-size scaling of the numerical lattice calculations close to the adiabatic limit, and comparison with the exact CFT result, for $k = 2$, $\lambda = 0.3$, and different system sizes $L$ ranging from 62 to 182 sites.
As in Fig.~\ref{introfig_loschmidt}, the agreement between numerics and analytical results is striking, now including the nonadiabatic region where $T/L\ll1$.%
}%
\label{Fig:Numerics}
\end{figure*}

Note that the ground state ($h=0$) driven by deformations \eqref{s435} with $k=1$ has $h_1=0$, owing to the definition above Eq.~\eqref{p0pt_SL2}.
As a result, the corresponding amplitude \eqref{p0pt_SL2} is a pure phase.
This is due to the fact that the ground state of a CFT is invariant under the `global conformal group' of deformations \eqref{s435} with $k=1$, as noted below Eqs.~\eqref{cG_cF_mn}.
(See also Footnote~\ref{foom}.)
It is the reason why our simulations of $\mathrm{SL}(2,\RR)$ drives in Figs.~\ref{introfig_loschmidt}(a) and \ref{Fig:Numerics} are carried out with $k=2$, as the latter is the lowest harmonic that gives rise to a nontrivial return probability for the ground state.
[Incidentally, $\mathrm{SL}(2,\RR)$ drives with $k=1$ provide a good numerical check of the validity of CFT as an effective description of gapless lattice systems at low energies;
see Appendix \ref{App:LattCalcs}.]

\paragraph{Adiabatic limit.}
As before, one is most interested in the return amplitude \eqref{s165} after one period, in the adiabatic regime $L/T \ll 1$.
Taking $t=T$ in Eq.~\eqref{p0pt_SL2} and expanding \eqref{eq:u_t} in that regime yields
\beq
\label{psizo}
\langle \psi(0)|\psi(T) \rangle
= e^{i \Theta} \Bigl[ 1 + O\bigl(L^2/T^2\bigr) \Bigr]
\eeq
for $\Theta = 2\pi \bigl(
- h_0 \frac{T}{L}
+ h_{k} \bigl[ \cosh(2\lambda) - 1 \bigr]
+ h_{k} \frac{\sinh^2(2\lambda)}{2} \frac{L}{T}
\bigr) + O(L^2/T^2)$.
The latter manifestly involves a dynamical phase, a Berry phase, a first nonadiabatic correction, and a rest term.
The Berry phase, in particular, is consistent with Eq.~\eqref{s43} for a rotating drive in $\mathrm{SL}(2,\RR)$ \cite{Oblak:2017}.

Note that the return amplitude \eqref{psizo} has unit norm up to $L^2/T^2$ corrections, so it is those corrections that are responsible for the nontrivial return probability $|\langle \psi(0)|\psi(T) \rangle|^2$.
As in Sec.~\ref{Sec:QGadiab}, the corrections are most conveniently found by expanding in powers of the adiabatic parameter \eqref{epss_exact} rather than $L/T$.
One finds
\begin{align}
& |\langle \psi(0)|\psi(T) \rangle|^2 \nonumber \\
& =
1 - \epss^2 \frac{2h_{k}}{k} \sinh^2(2\lambda) \sin^2(\pi k/\epss) + O\bigl(\epss^3\bigr),
\label{echo}
\end{align}
which shows that there is an oscillating tail in the return probability beyond the adiabatic limit.
This result is a special case of the general adiabatic expansion \eqref{cL_Vir_final}.
A single Fourier mode $k$ now appears in the quantum metric \eqref{metryc}, since in fact the vector field $\bY_t = Y_t(\phii)\partial_{\phii}$ of Eq.~\eqref{Fpert} is explicitly given by $Y_t(\phii) = \sinh(2\lambda) \bigl[ \sin(k\phii) - \sin(k[\phii+\Omega t]) \bigr]$.
The oscillations of Eq.~\eqref{echo} with $T$ can then be understood by referring to Fig.~\ref{Fig:Hyperbolicdisk}: the micromotion around the leading adiabatic approximation has a (high) frequency $\Omega$ of its own, which may or may not resonate with the (low) frequency $\omega$ of the drive.
At resonance, when $\Omega T\in 2\pi\ZZ$, the initial and final rays coincide.

As at the end of Sec.~\ref{Sec:QGadiab}, one should keep in mind that the driven Hamiltonian of a nonchiral CFT involves both right and left movers, as in Eq.~\eqref{fullh}.
One should thus add to Eq.~\eqref{Ht_SL2} the same expression with $\stt$ replaced by $\overline\stt$.
This left-moving term only involves the generators $\{ \bar{L}_{-k}, \bar{L}_0, \bar{L}_k \}$, which similarly generate an $\mathfrak{sl}(2,\RR)$ Lie algebra that commutes with the right-moving one, so that a matrix representation can again be used to solve the Schr\"odinger equation.
In particular, the left-moving frequency $\overline\Omega$ in Eq.~\eqref{bb21b} becomes
\beq
\overline\Omega
=
\frac{2\pi}{L} \sqrt{ 1 + 2\cosh(2\lambda) (L/T) + (L/T)^2 },
\label{ttt25b}
\eeq
which is obtained from Eq.~\eqref{Omega_SL2} by changing the sign in front of $\omega=2\pi/T$ wherever it appears.
A key consequence is that the full nonchiral return probability \eqref{ss195bis}--\eqref{sss} is the product of \eqref{echo} with its left-moving counterpart:
\begin{align}
& |\langle \psi(0)|\psi(T) \rangle|^2
\label{p0pt_SL2_nonChir} \\
& \quad \begin{aligned}
= 1
& - \epss^2 \frac{2h_{k}}{k} \sinh^2(2\lambda) \sin^2(\pi k/\epss) \\
& - \bar\epss^2 \frac{2\bar h_{k}}{k} \sinh^2(2\lambda) \sin^2(\pi k/\bar\epss)
+ O\bigl(\epss^3,\bar\epss^3\bigr),
\end{aligned} \nonumber
\end{align}
where $\epss \equiv \omega/\Omega \neq \omega/\overline{\Omega} \equiv \bar\epss$ and $\bar h_k \equiv \bar h + \tfrac{\bar c}{24}(k^2-1)$.
This is the nonchiral version of the chiral result \eqref{echo}.
It is plotted (with a dashed red curve) in Fig.~\ref{introfig_loschmidt}(a), and displays a characteristic oscillating convergence as $T/L\to\infty$.
However, in contrast to the purely chiral case \eqref{echo}, it never quite reaches its asymptotic value due to the mismatch $\Omega \neq \overline{\Omega}$ [compare Eqs.~\eqref{Omega_SL2} and \eqref{ttt25b}].

\paragraph{Comparison with lattice results.}
The comparison between CFT and a lattice model is displayed in Fig.~\ref{Fig:Numerics}.
There, we plot return probabilities for a periodic free-fermion chain at half filling with $N$ sites and driven inhomogeneous hoppings
\beq
\label{Jj_t_SL2}
J_{j}(t)
= \cosh(2\lambda) - \sinh(2\lambda)\cos(k[2\pi j/N - \omega t]),
\eeq
corresponding to the spin chain \eqref{eq:H_XXZ_t} for $\Delta = 0$ with the same nearest-neighbor couplings.
The return probability $|\langle\psi_{N}(0)|\psi_{N}(T)\rangle|^2$ of the lattice quantum state is computed numerically by approximating the evolution operator with a finite number of steps and using exact diagonalization for each time step; see Appendix~\ref{App:LattCalcs}.
The result exhibits remarkable agreement with the analytical formula obtained by taking the squared complex norm of Eq.~\eqref{p0pt_SL2}, and multiplying it by itself with $\omega$ replaced by $-\omega$ everywhere (including in $\Omega$).
This is true even far away from the adiabatic regime, as long as the strength of the perturbation (controlled by $\lambda$) is sufficiently weak.
Importantly, CFT results for the return probability
are even applicable to lattice systems with a relatively small number of lattice sites, as shown in Fig.~\ref{Fig:Numerics}(b).
This suggests that our dynamical probes for quantum geometry can realistically be implemented in quantum simulators with fewer than a hundred sites.

%==========================================================
\section{Discussion}
\label{Sec:Disc}
%==========================================================

This work explored the infinite-dimensional quantum geometry of 1+1D gapless many-body quantum systems through the lens of their low-energy dynamics, modeled as driven conformal field theories (CFTs).
For slowly driven systems, we showed that the return probability probes the quantum metric through oscillations as a function of the driving frequency; see Figs.~\ref{introfig_loschmidt} and~\ref{Fig:Numerics}.
In the complementary limit of arbitrarily fast but weak drives, absorption rates and linear-response coefficients respectively give access to the metric and the Berry curvature of the same quantum geometry.

These results are universal in the sense that they apply to any 1+1D gapless quantum system with an emergent conformal symmetry at low energies.
They hold for both free and strongly interacting systems, and only depend on universal parameters such as the central charge of the effective theory.
This is supported by a remarkable agreement between our field-theoretic results and numerics for gapless lattice models, opening a path to observe Virasoro quantum geometry in state-of-the-art quantum simulators.

Promising future research directions include:

\textit{Geometric formulation of Floquet drives}.
It was recently understood that Floquet theory can be formulated from a quantum geometric perspective \cite{schindler2025geometricfloquettheory}, by decomposing the Floquet evolution operator into a purely dynamical part (the `average-energy operator') and a geometric part.
It would be interesting to compute the average-energy operator for our periodic drives, and understand the link between the Virasoro Berry phases \eqref{s43} \cite{Oblak:2017} and those of \cite{schindler2025geometricfloquettheory}.

\textit{Higher dimensions}.
Our study of the quantum geometry of 1+1D gapless many-body quantum systems relied on the Virasoro algebra, allowing us to probe an infinite-dimensional parameter space \eqref{s125q}.
While this algebraic structure is absent in higher-dimensional CFTs, one may still design nonequilibrium probes of the quantum geometry associated with the global conformal group SO$(d+1,2)$ of gapless many-body systems in $d$+1 spacetime dimensions.
Indeed, given that the approach of this work holds for any continuous symmetry group, it could naturally be applied to higher-dimensional CFTs.

\textit{Quantum geometry of mixed states}.
Our work concerned nonequilibrium probes of the quantum geometry for pure states, specifically Virasoro coherent states \eqref{e218}.
The relevant quantum metric then was (the pullback of) the Fubini-Study metric.
A natural next step is to consider mixed states, and dynamically probe their quantum geometry.
In the strict adiabatic regime, one may expect a suitable generalization of our protocol to feature the Uhlmann phase \cite{Uhlmann1986ParallelTA, Uhlmann:1990er}, as a mixed-state counterpart of the Berry phase.
Crucially though, the corresponding metric is not unique \cite{BengtssonZyczkowski:2006}, so subleading corrections in the adiabatic regime would probe different metrics depending on the choice of distance measure.
Examples include the Bures metric and the Kubo-Mori metric; the former was recently considered in \cite{JiEtAl:2025, GuanBradlyn:2025}, and the latter was studied in \cite{deBoerEtAl:2023} for thermal states in driven CFT.

\paragraph{Acknowledgments.}
We are especially grateful to Bruno Mera for useful comments on the first version of this paper.
We also thank Mathieu Beauvillain, Marin Bukov, Adolfo del Campo, Rapha\"el Ducatez, Benoit Estienne, Gian Michele Graf, Gregor Jotzu, Shinsei Ryu, and Erik Tonni for inspiring discussions.
B.L.\ acknowledges financial support from the Swiss National Science Foundation (Postdoc.Mobility Grant No.~214461).
P.M.\ acknowledges financial support from the Wenner-Gren Foundations (Grant No.\ FT2022-0002).

%==========================================================
\appendix
%==========================================================

%==========================================================
\section{Brief dictionary of CFT formulas}
\label{App:CFTdict}
%==========================================================

\newcommand\Tstrut{\rule{0pt}{3.0ex}}       % top strut
\newcommand\Bstrut{\rule[-1.5ex]{0pt}{0pt}} % bottom strut

\begin{table*}[!ht]
\centering
\begin{tabular}{c|l|l|}
& Angle coordinates
& Length coordinates \\
\hline
\Tstrut
{\small \multirow{2}{*}{$\Diff(S^1)$} }
& {\small $f(\phii)$ with $f'(\phii) > 0$}
& {\small $f(x) = \frac{L}{2\pi} f(\phii)$ with $\partial_{x}f(x) > 0$} \\
& {\small $f(\phii + 2\pi) = f(\phii) + 2\pi$}
& {\small $f(x + L) = f(x) + L$} \Bstrut \\
\hline
\Tstrut
{\small \multirow{6}{*}{$\Vect(S^1)$} }
& {\small $\bX = X(\phii) \partial_{\phii}$,\; $X(\phii + 2\pi) = X(\phii)$}
& {\small $\bX = X(x) \partial_{x}$,\;  $X(x + L) = X(x)$} \Bstrut \\
\Tstrut
& {\small $X(\phii)
=  \underset{n\in\ZZ}{\sum} e^{i n \phii} \hat{X}_{n}$}
& {\small $X(x)
= \frac{L}{2\pi} X(\phii)
= \frac{L}{2\pi}  \underset{n\in\ZZ}{\sum} e^{i \frac{2\pi}{L}nx} \hat{X}_{n}$} \\
& \multicolumn{2}{l|}{{\small $\hat{X}_{m} = \frac{1}{2\pi} \oint \dd \phii\, e^{-i m \phii} X(\phii)$, \; $X(\phii) \in \RR$ $\iff$ $\overline{\hat{X}_{m}} = \hat{X}_{-m}$}} \Bstrut \\
\Tstrut
& \multicolumn{2}{l|}{{\small $[\bX, \bY]
= \bigl[ X(\phii) Y'(\phii) - Y(\phii) X'(\phii) \bigr] \partial_{\phii} = \bigl[ X(x) \partial_{x}Y(x) - Y(x) \partial_{x}X(x) \bigr] \partial_{x}$}} \Bstrut \\
\Tstrut
& {\small Infinitesimal $f(\phii) = \phii + \epsilon X(\phii)$}
& {\small Infinitesimal $f(x) = x + \epsilon X(x)$} \Bstrut \\
\hline
\Tstrut
{\small \multirow{4}{*}{\begin{tabular}{@{}c@{}}Virasoro \\ algebra \end{tabular}} }
& {\small $\stt_{\pm}(\phii) = \frac{1}{2\pi} \underset{n\in\ZZ}{\sum} e^{\pm i n \phii} \bigl( L^{\pm}_{n} - \frac{c}{24} \delta_{n,0} \bigr)$}
& {\small $\stt_{\pm}(x) = \bigl( \frac{2\pi}{L} \bigr)^2 \stt_{\pm}(\phii) = \frac{2\pi}{L^2} \underset{n\in\ZZ}{\sum} e^{\pm i \frac{2\pi}{L} n x} \bigl( L^{\pm}_{n} - \frac{c}{24} \delta_{n,0} \bigr)$} \\
& \multicolumn{2}{l|}{{\small $L^{\pm}_{m} = \oint \dd \phii\, e^{\mp i m \phii}\, \stt_{\pm}(\phii) + \frac{c}{24} \delta_{m,0}$ satisfying Eq.~\eqref{e24}}} \Bstrut \\
\Tstrut
& {\small $\bigl[ \stt_{\pm}(\phii), \stt_{\pm}(\theta) \bigr] = \mp 2i \delta'(\phii-\theta) \stt_{\pm}(\theta)$} \hspace{4.2mm}
& {\small $\bigl[ \stt_{\pm}(x), \stt_{\pm}(y) \bigr] = \mp 2i \partial_{x} \delta(x-y) \stt_{\pm}(y)$} \\
& \; {\small$\pm i \delta(\phii-\theta) \stt_{\pm}'(\theta) \pm \frac{c}{24\pi} i \delta'''(\phii-\theta)$}
& \; {\small $\pm i \delta(x-y) \partial_{y}\stt_{\pm}(y) \pm \frac{c}{24\pi} i \partial_{x}^3 \delta(x-y)$} \Bstrut \\
\hline
\Tstrut
{\small \multirow{3}{*}{\begin{tabular}{@{}c@{}}Proj.\ anti-\\ Hermitian \\ rep.\ $\cu_{\pm}(\bX)$ \end{tabular}} }
& \multicolumn{2}{l|}{{\small$\cu_{\pm}(\bX)$ for $\bX \in \Vect(S^1)$ or $\Vect(S^1) \otimes \CC$ satisfying Eq.~\eqref{s95}}} \Bstrut \\
\Tstrut
& \multicolumn{2}{l|}{{\small $\cu_{\pm}(\bX)
= \partial_{\epsilon} \cU_{\pm}(\II + \epsilon \bX) \big|_{\epsilon=0}
= \mp i \oint \dd \phii\, X(\phii) \stt_{\pm}(\phii)
= \mp i \underset{m\in\ZZ}{\sum} \hat{X}_{m} \bigl( L^{\pm}_{\mp m} - \frac{c}{24} \delta_{m,0} \bigr)
= \mp i \int_{0}^{L} \dd x\, X(x) \stt_{\pm}(x)$}} \\
& \multicolumn{2}{l|}{{\small $L^{\pm}_{n}
= - \cu_{\pm}(\ell_{\mp n}) + \frac{c}{24} \delta_{n,0}$ for $\ell_n = -i e^{-in \phii} \partial_\phii$ satisfying $[\ell_{m}, \ell_{n}] = (m-n) \ell_{m+n}$}} \Bstrut \\
\hline
\Tstrut
{\small \multirow{2}{*}{\begin{tabular}{@{}c@{}}Proj.\ unitary \\ rep.\ $\cU_{\pm}(f)$ \end{tabular}} }
& \multicolumn{2}{l|}{{\small $\cU_{\pm}(f)$ for $f \in \Diff(S^1)$ satisfying 
$\cU_{\pm}(f) \cU_{\pm}(g)
= e^{\pm i c \hspace{0.7pt}\sfC(f,g)} \cU_{\pm}(f \circ g)$ with $\sfC(f,g)$ in Eq.~\eqref{bott}}}
\Bstrut \\
\Tstrut
& \multicolumn{2}{l|}{{\small $\cU_{\pm}\bigl(e^{\epsilon \bX}\bigr) = e^{\epsilon \cu_{\pm}(\bX)}$ for $\bX \in \Vect(S^1)$, $\epsilon \in \RR$ $\implies$ $\cU_{\pm}(\II + \epsilon \bX) = 1 + \epsilon \cu_{\pm}(\bX) + o(\epsilon)$}} \Bstrut \\
\hline
\end{tabular}
\caption{%
Useful formulas for nonchiral 1+1D CFT in angle coordinates and their counterparts in length coordinates.%
}%
\label{Table:CFT_dictionary}
\end{table*}

This appendix complements the concise review of CFT and Virasoro symmetry in Sec.~\ref{Sec:VirQG}, extended to the nonchiral case of Eq.~\eqref{fullh}.
For convenience, we change notation to subscripted $+$ or $-$ to respectively indicate right or left movers.
This includes \eg $\stt_{+}(\phii^{-}) = \stt(\phii^{-})$ and $\stt_{-}(\phii^{+}) = \overline{\stt}(\phii^{+})$, each depending only on its respective light-cone coordinate $\phii^{\pm}$ defined above Eq.~\eqref{cot}.
However, we set $c = \bar{c}$ for simplicity.

Table~\ref{Table:CFT_dictionary} summarizes useful formulas for nonchiral 1+1D CFT
in angle coordinates $\phii$, $\theta$ vs.\ length coordinates $x \equiv \frac{L}{2\pi} \phii$, $y \equiv \frac{L}{2\pi} \theta$.
Quantities are distinguished by their arguments, $\delta(\phii)$ and $\delta(x)$ respectively denote the $2\pi$- and $L$-periodic delta functions, a prime ${}'$ denotes differentiation with respect to $\phii$, and $\oint \dd\phi$ is shorthand for $\int_{0}^{2\pi} \dd\phi$.
Moreover, $\Diff(S^1)$ denotes the universal cover of the group of orientation-preserving diffeomorphisms of the circle $S^1$.
$\Vect(S^1)$ denotes the corresponding Lie algebra of real, smooth vector fields on $S^1$, and $\Vect(S^1) \otimes \CC$ denotes its complexification.

Some context and details for the algebra representation $\cu_{\pm}(\bX)$ in Table~\ref{Table:CFT_dictionary} are missing.
To fill this gap, first note that Eq.~\eqref{e29} can be written as
\beq
\bX
=
\sum_{m\in\ZZ}
\hat{X}_{m}\,i\ell_{-m}
\quad\text{with}\quad
\ell_m \equiv -ie^{-im\phii}\der_{\phii}
\eeq
in terms of complexified generators $\ell_{m}$ satisfying the commutation relations $[\ell_{m}, \ell_{n}] = (m-n) \ell_{m+n}$ of the Witt algebra.\footnote{This is the subalgebra $\operatorname{span}_{\CC} (\{ \ell_n \}_{n\in\ZZ})$ of the complexification $\Vect(S^1) \otimes \CC$. Note the isomorphism of the Witt algebra under $\ell_{n} \mapsto - \ell_{-n}$, which does not carry over to the Virasoro algebra \eqref{e24} since the central term changes sign.}
The table's nonchiral version of Eq.~\eqref{cuX_Lm} thus says that $L^{\pm}_m \equiv \cu_{\pm} \bigl( -\ell_{\mp m} \bigr) + \frac{c}{24} \delta_{m,0}$ represents the (complexified) vector field $\ell_m$.
Now, owing to the table's nonchiral version of the projective composition law \eqref{UfUg_Vir}, commutators of the operators $\cu_{\pm}(\bX)$ extend the algebra $\Vect(S^1)$ by a central term:
\beq
\label{s95}
\big[\cu_{\pm}(\bX),\cu_{\pm}(\bY)\big]
=
\cu_{\pm}\big({-}[\bX,\bY]\big) \pm i c\hspace{1.8pt}\sfc(\bX, \bY)
\eeq
with $\sfc(\bX, \bY)
\equiv \oint \frac{\dd \phii}{24\pi} X'(\phii) Y''(\phii)$.
The latter is the so-called Gelfand-Fuchs cocycle \cite{GuieuRoger:2007, KhesinWendt:2009}---the algebra counterpart of the Bott cocycle \eqref{bott}.
In fact, all nontrivial central extensions of $\Vect(S^1)$ by $\RR$ are isomorphic and generated by $\sfc(\cdot, \cdot)$ (see \eg \cite{KhesinWendt:2009}), so the Virasoro central extension is unique up to redefinitions of the generators.
For the light-cone components $\stt_{\pm}(\theta) = i \cu_{\pm}(\delta(\phii-\theta)\partial_{\phii})$ of the stress-energy tensor, Eq.~\eqref{s95} yields
\begin{multline}
\bigl[ \stt_{\pm}(\phii), \stt_{\pm}(\theta) \bigr]
= \mp 2i \delta'(\phii-\theta)\, \stt_{\pm}(\theta) \\
  \pm i \delta(\phii-\theta)\, \stt_{\pm}'(\theta)
  \pm \frac{c}{24\pi}\, i\, \delta'''(\phii-\theta),
\label{e24_pos_space}
\end{multline}
which are nothing but the Virasoro commutation relations \eqref{e24} upon using
the nonchiral version of Eq.~\eqref{e28} in the table.
Lastly, note that Hermiticity $(L^{\pm}_m)^{\dagger} = L^{\pm}_{-m}$ follows directly from the unitarity of $\cU_{\pm}(\cdot)$, or equivalently from the anti-Hermiticity of $\cu_{\pm}(\cdot)$.

%==========================================================
\section{Quantum geometry away from identity}
\label{App:VirQGnonId}
%==========================================================

In this appendix, we briefly describe the quantum metric and the Berry curvature of a homogeneous parameter space \eqref{s34} at points away from the identity.
We then apply this to the case \eqref{s125q} that is relevant for Virasoro quantum geometry.

\paragraph{General case.}
Let $G$ be a Lie group, $\cU$ a unitary representation of $G$, and consider the parameter space \eqref{s34} for `deformed' Hamiltonians of the form $\cU(f)H_0\cU(f)^{-1}$.
We saw in Sec.~\ref{Sec:VirQG} that the ensuing quantum metric and Berry curvature at the identity are given by Eq.~\eqref{tb18}.
When evaluated at the coset $fG_0$ in the space \eqref{s34}, the metric and the curvature are given by 
\begin{subequations}
\label{GFF}
\begin{align}
\label{ss18}
\cG
& =
- \langle h|\cu(f^{-1}\dd f) \odot\cu(f^{-1}\dd f)|h\rangle \nonumber \\
& \quad\, + \langle h|\cu(f^{-1}\dd f)|h\rangle \odot \langle h|\cu(f^{-1}\dd f)|h\rangle, \\
\label{tt18}
\cF
& =
2i\langle h| \cu(f^{-1}\dd f) \wedge \cu(f^{-1}\dd f) |h\rangle,
\end{align}
\end{subequations}
where $\odot$ and $\wedge$ are defined as in Eq.~\eqref{e220}, and $\cu(f^{-1}\dd f)$ is the representation of the left Maurer-Cartan form $f^{-1}\dd f$ on $G$.
The latter is the map $f^{-1}\dd f:T_fG\to\mathfrak{g}$ such that, for any path $\gamma(t)$ in $G$ for which $\gamma(0)=f$, one has
\beq
f^{-1}\dd f(\der_t\gamma|_{t=0})
\equiv
\der_t\big(f^{-1}\cdot\gamma(t)\big) \big|_{t=0}
\label{s115}
\eeq
with the dot $\cdot$ denoting multiplication in $G$.

\paragraph{Virasoro case.}
For $G=\Diff(S^1)$, the metric and the curvature at the identity are given by Eqs.~\eqref{cG_cF_XmYn}--\eqref{cG_XY}.
Let us now extend these expressions to the whole parameter space \eqref{s125q}.
Owing to Eq.~\eqref{GFF}, this is essentially achieved by replacing $\bX,\bY$ in Eqs.~\eqref{cF_XY}--\eqref{cG_XY} by the left Maurer-Cartan form $f^{-1}\dd f$, and including tensor products where and when they are needed.
Let us first discuss the Maurer-Cartan form in its own right.
As briefly explained below Eq.~\eqref{GFF}, $f^{-1}\dd f$ is a Lie-algebra-valued one-form; so, for $G=\Diff(S^1)$, it takes values in the space $\mathfrak{g}=\Vect(S^1)$ of vector fields on the circle.
Any tangent vector at $f\in\Diff(S^1)$ can be seen as the time derivative $\der_t\gamma_t|_{t=0}$ of a path $\gamma_t$ in $\Diff(S^1)$ such that $\gamma_0=f$.
This allows us to write the left Maurer-Cartan form on $\Diff(S^1)$ by adapting Eq.~\eqref{s115} to a group operation given by the composition of functions:
\beq
f^{-1}\dd f\big(\der_t\gamma_t |_{t=0}\big)
\equiv
\der_t \bigl( f^{-1} \circ \gamma_t \bigr) \big|_{t=0}
=
\frac{\der_t\gamma_t(\phii)|_{t=0}}{f'(\phii)}\der_\phii.
\label{s12t}
\eeq
Here we used the chain rule to obtain the last equality.
As expected, the Maurer-Cartan form is vector-field-valued.
Henceforth, we write it as
\beq
f^{-1}\dd f
\equiv
\frac{\delta f(\phii)}{f'(\phii)}\der_{\phii},
\label{s12q}
\eeq
with the functional differential $\delta f$ emphasizing that the one-form acts on a space of functions.

The compact notation \eqref{s12q} can now be combined with Eqs.~\eqref{cF_XY}--\eqref{cG_XY} to express the quantum metric and the Berry curvature at any point $f$ in the group manifold $\Diff(S^1)$ [or more precisely at any coset $f\circ S^1$ in the parameter space \eqref{s125q}].
Namely,
\begin{widetext}
\begin{subequations}
\begin{align}
\cG
& =
\oint\frac{\dd\phii\,\dd\theta}{16\pi^2}
\frac{(h{-}\frac{c}{24})
\Big(\left.\frac{\delta f}{f'}\right|_{\phii}{-}\left.\frac{\delta f}{f'}\right|_{\theta}\Big)^2
+
\frac{c}{24}
\Big(\Big(\left.\frac{\delta f}{f'}\right|_{\phii}{-}\left.\frac{\delta f}{f'}\right|_{\theta}\Big)\!\Big.'\,\Big)^2}{\sin^2[(\theta-\phii)/2]}, \\[.7em]
\label{ffa}
\cF
& =
-2\oint\frac{\dd\phii}{2\pi}\Big[\Big(h-\frac{c}{24}\Big)\frac{\delta f}{f'}\wedge\Big(\frac{\delta f}{f'}\Big)'
-\frac{c}{24}\frac{\delta f}{f'}\wedge\Big(\frac{\delta f}{f'}\Big)'''\,\Big],
\end{align}
\end{subequations}
where we use the notation $(\delta f)^2 \equiv \delta f \odot \delta f$ for symmetrized products of one-forms.
These expressions are explicit in principle, but complicated to use in practice.
Again, the main lesson is that both of these tensor fields are fully determined by their value at the identity.
In particular, we saw in Sec.~\ref{Sec:QGadiab} that the return probability of a slowly driven quantum state involves the quantum metric evaluated at a point $f$ far away from the identity, even though the actual formula involves the much simpler metric in Eqs.~\eqref{cG_XmYn} or \eqref{cG_XY} at the identity.

%==========================================================
\section{Linear response to Virasoro drives}
\label{App:LinResp}
%==========================================================

Our goal here is to derive the linear-response result \eqref{htcOht_suscep} and its gradient expansion \eqref{htcOht_grad_exp}.
The first step is to write the Hamiltonian \eqref{Ht_uXt} as
$H(t) = H_0 + V_{\textrm{S}}(t) + O(\epsilon^2)$ with
\beq
V_{\textrm{S}}(t)
\equiv
\epsilon \bigl[ \cu(\bX_t), H_{0} \bigr]
= i \epsilon \frac{2\pi}{L} \sum_{n \neq 0} n \hat{X}_{n}(t) L_{-n},
\eeq
where $\bX_t$ is an arbitrary time-dependent vector field.
As in the main text, let $|\phi(0)\rangle = |h \rangle$ be the initial state and consider the solution $|\phi(t)\rangle$ of the Schr\"odinger equation at time $t$, under the full $H(t)$ dynamics.

\paragraph{Susceptibility.}
Since we wish to compute $\langle \phi(t)| i\cu(\bW) | \phi(t)\rangle$ for any vector field $\bW$, it suffices to study the single-mode expectation value $\langle \phi(t)| L_{-m} | \phi(t)\rangle$ for any given integer $m$.
[Indeed, $i\cu(\bW)$ is a linear combination of $L_m$\hspace{0.5pt}s in Eq.~\eqref{cuX_Lm}.]
To this end, we work in the interaction picture and define
\beq
V_{\textrm{I}}(t)
\equiv e^{i H_0 t} V_{\textrm{S}}(t) e^{-i H_0 t} \\
= i \epsilon \frac{2\pi}{L} \sum_{n \neq 0} n \hat{X}_{n}(t) L_{-n}^{\textrm{I}}(t) \\
= i \epsilon \frac{2\pi}{L} \sum_{n \neq 0} n \hat{X}_{n}(t)\,
L_{-n} e^{i 2\pi nt/L}
\eeq
with
$L_{m}^{\textrm{I}}(t)
\equiv e^{i H_0 t} L_{m} e^{-i H_0 t}
= L_m e^{-i 2\pi mt/L}$
obtained by solving
$\partial_t L_{m}^{\textrm{I}}(t)
= - i \frac{2\pi}{L} m L_{m}^{\textrm{I}}(t)$
with $L_{m}^{\textrm{I}}(0) = L_{m}$.
Thus, $\langle \phi(t)| L_{-m} |\phi(t)\rangle = \langle h| U_{\textrm{I}}(t)^{-1} L_{-m}^{\textrm{I}}(t) U_{\textrm{I}}(t) |h \rangle$, where the evolution operator $U_{\textrm{I}}(t)$ is expressible using the Dyson formula $U_{\textrm{I}}(t) = I - i \int_0^t \dd s\, V_{\textrm{I}}(s) + O(\epsilon^2)$.
This gives the Dyson series
\begin{align}
\langle \phi(t)| L_{-m} |\phi(t)\rangle
& = \langle h| L_{-m}^{\textrm{I}}(t) |h \rangle
    + i \int_0^t \dd s\, \langle h| \bigl[ V_{\textrm{I}}(s), L_{-m}^{\textrm{I}}(t) \bigr] |h \rangle
    + O(\epsilon^2) \nonumber \\
& = \langle h| L_{-m} |h \rangle e^{i 2\pi mt/L}
    + i\epsilon \frac{2\pi}{L} \sum_{n \neq 0} \int_0^t \dd s\, n\hat{X}_{n}(s) \cF_{m,n} e^{i 2\pi (mt + ns)/L}
    + O(\epsilon^2) \nonumber \\
& = \langle h| L_{-m} |h \rangle e^{i 2\pi mt/L}
    + \epsilon \sum_{n \neq 0} \int_0^t \dd s\, \chi_{m,n}(t,s) \hat{X}_{n}(s)
    + O(\epsilon^2),
\label{htLmht_suscep}
\end{align}
where we identified the Berry curvature $\cF_{m,n}$ of Eq.~\eqref{cF_mn} and the susceptibility $\chi_{m,n}(t,s)$ of Eq.~\eqref{chi_mn}.
For operators $i\cu(\bW)$ with Fourier components $\hat{W}_{m}^{\phantom{*}} = \hat{W}_{-m}^*$, the result \eqref{htcOht_suscep} follows.

\paragraph{Gradient expansion.}
It remains to derive the gradient expansion \eqref{htcOht_grad_exp}.
To this end, we use repeated integration by parts starting from Eq.~\eqref{htLmht_suscep}, which yields
\begin{align}
\langle \phi(t)| L_{-m} |\phi(t)\rangle
& = \langle h| L_{-m} |h \rangle e^{i 2\pi mt/L}
    + \epsilon \sum_{n \neq 0} \cF_{m,n} \biggl[ \hat{X}_{n}(s) e^{i 2\pi (mt + ns)/L} \biggr]_{s=0}^{t} \\
& \quad + \epsilon \sum_{n \neq 0} \cF_{m,n} \sum_{k=1}^{N} \biggl( - \frac{L}{2\pi i n} \biggr)^k \biggl[ \bigl[\partial_{s}^k \hat{X}_{n}(s)\bigr] e^{i 2\pi (mt + ns)/L} \biggr]_{s=0}^{t}
    + R_m^{(N)}(t) + O(\epsilon^2) \nonumber
\end{align}
for $N \geq 1$, with the remainder term
\beq
R_m^{(N)}(t)
= - \epsilon \sum_{n \neq 0} \cF_{m,n} \biggl( - \frac{L}{2\pi i n} \biggr)^{N} \int_0^t \dd s\, [\partial_{s}^{N+1} \hat{X}_{n}(s)] e^{i 2\pi (mt + ns)/L}.
\eeq
Since $\cF_{m,n}$ is zero if $m = 0$ (or $n = 0$), it follows from the above that
\begin{align}
& \langle \phi(t)| i\cu(\bW) |\phi(t)\rangle
= 
\langle h| i\cu(\bW) |h \rangle e^{i 2\pi mt/L}
  + \epsilon \sum_{m,n \neq 0} \hat{W}_{m} \cF_{m,n}
    \left[ \hat{X}_{n}(s) e^{i 2\pi (mt + ns)/L} \right]_{s=0}^{t} \nonumber \\
& \qquad
  + \epsilon \sum_{m,n \neq 0} \hat{W}_{m} \cF_{m,n} \sum_{k=1}^{N} \biggl( - \frac{L}{2\pi i n} \biggr)^k \biggl[ \bigl[\partial_{s}^k \hat{X}_{n}(s)\bigr] e^{i 2\pi (mt + ns)/L} \biggr]_{s=0}^{t} + \sum_{m \neq 0} \hat{W}_{m} R_m^{(N)}(t)
  + O(\epsilon^2).
\end{align}
This implies Eq.~\eqref{htcOht_grad_exp} since $\cF_{m,n}$ is zero unless $m + n = 0$.

%==========================================================
\section{Rotating \texorpdfstring{$\mathrm{\bf SL}\boldsymbol(\boldsymbol2\boldsymbol,\mathbb{R}\boldsymbol)$}{SL(2,R)} drives}
\label{App:SL2R}
%==========================================================

This appendix completes Sec.~\ref{Sec:SL2R}, for which it provides the explicit matrix expressions mentioned there.

\paragraph{Hamiltonian as a matrix.}
Our starting point is the Hamiltonian \eqref{eq:Ht_SL2_Ls} expressed as
\begin{align}
\label{eq:Ht_SL2_Ks}
H(t)
= \frac{2\pi k}{L} \biggl[ \cosh(2\lambda) K_0
  + \frac{i}{2} \sinh(2\lambda)
    \biggl( e^{ik\omega t} K_- - e^{-ik\omega t} K_+ \biggr) \biggr]
  - \frac{\pi c}{12L} k^2
\end{align}
in terms of the following generators of $\mathfrak{su}(1,1) \cong \mathfrak{sl}(2,\RR)$ \cite{Perelomov:1977, Perelomov:1986, ScullyZubairy:1997}:
\beq
\label{su11}
K_{0}
\equiv
\frac{1}{k} \Bigl[ L_{0} + \frac{c}{24} (k^2-1) \Bigr],
\qquad
K_{\mp}
\equiv
\pm \frac{i}{k} L_{\pm k}.
\eeq
These satisfy the usual $\mathfrak{su}(1,1)$ relations
\beq
\left[ K_{-}, K_{+} \right] = 2K_{0},
\qquad
\left[ K_{0}, K_{\pm} \right] = \pm K_{\pm},
\qquad
K_{0}^\dagger = K_{0},
\qquad
K_{-}^\dagger = K_{+}.
\eeq
By convention, the generators $K_{\pm}$ have a factor $i$, meaning that the corresponding coefficients $\beta_{\pm} = \mp \frac{i \pi k}{L} \sinh(2\lambda) e^{\mp ik\omega t}$ in Eq.~\eqref{eq:Ht_SL2_Ks} satisfy $\overline{\beta_{+}} = \beta_{-}$.

To make analytical progress, we use the fundamental representation of $\mathfrak{su}(1,1)$ in terms of $2\times2$-matrices:
\beq
\begin{gathered}
\label{2x2_rep}
K_{0} \big|_{2\times2}
=
\begin{pmatrix} -1/2 & 0 \\ 0 & 1/2 \end{pmatrix},
\qquad
K_{-} \big|_{2\times2}
= \begin{pmatrix} 0 & 1 \\ 0 & 0 \end{pmatrix},
\qquad
K_{+} \big|_{2\times2}
= \begin{pmatrix} 0 & 0 \\ -1 & 0 \end{pmatrix}.
\end{gathered}
\eeq
The Hamiltonian \eqref{eq:Ht_SL2_Ks} in this picture reads
\begin{align}
\label{Ht_SL2_2x2}
H(t)\big|_{2\times2}
= \frac{\pi k}{L}
  \begin{pmatrix}
    -\cosh(2\lambda) - ck/12 & i \sinh(2\lambda) e^{ik\omega t} \\
    i \sinh(2\lambda) e^{-ik\omega t} & \cosh(2\lambda) - ck/12
  \end{pmatrix}.
\end{align}
Note that the fundamental representation is nonunitary since 
$\mathrm{SU}$(1,1) is noncompact---as is manifest in the non-Hermitian matrix \eqref{Ht_SL2_2x2}.
However, the fundamental representation is faithful, so any lesson learned in this representation holds for $\mathfrak{su}(1,1)$ in general.

\paragraph{Evolution operator.}
By the definition of time-ordered exponentials as product integrals, the time-evolution operator \eqref{Ut_def} can be represented as in Eq.~\eqref{Ut_Trotter}.
The matrix representation \eqref{Ht_SL2_2x2} of the Hamiltonian thus yields
\beq
\label{Ut_SL2_2x2}
U(t) \big|_{2\times2}
= e^{i \frac{\pi c}{12L}k^2 t}
  \lim_{M\to\infty} \prod_{m=0,\ldots,M}^{\longleftarrow}
  \exp \biggl[ \frac{\pi k}{L} \delta t \begin{pmatrix}
    i \cosh(2\lambda) & \sinh(2\lambda) e^{ik\omega m \delta t} \\
    \sinh(2\lambda) e^{-ik\omega m \delta t} & -i \cosh(2\lambda)
  \end{pmatrix} \biggr],
\eeq
where $\delta t\equiv t/M$ and the product $\overset{\longleftarrow}{\prod}$ is ordered so that time increases from right to left.
Noting that all the dependence on `time' $m$ is contained in the phases $e^{\pm i k\omega m\delta t}$, we rewrite each factor in Eq.~\eqref{Ut_SL2_2x2} as
\beq
\begin{pmatrix}
  0 & e^{-i\omega s/2} \\
  e^{i\omega s/2} & 0
\end{pmatrix}
\begin{pmatrix}
  -i\bar{\alpha} & \bar{\beta} \\
  \beta & i\alpha
\end{pmatrix}
\begin{pmatrix}
  0 & e^{-i\omega s/2} \\
  e^{i\omega s/2} & 0
\end{pmatrix}
=
\begin{pmatrix}
  i\alpha & \beta\,e^{-i\omega s}\\
  \bar{\beta}\,e^{i\omega t} & -i\bar{\alpha} 
\end{pmatrix}
\eeq
for $s = -km \delta t$ and $\alpha = \cosh(2\lambda)$, $\beta = \sinh(2\lambda)$.
It follows that each pair of consecutive factors in the product \eqref{Ut_SL2_2x2} gives rise to products of the form
\beq
\begin{pmatrix}
  0 & e^{i k\omega (m+1)\delta t/2} \\
  e^{-i k\omega (m+1)\delta t/2} & 0
\end{pmatrix}
\begin{pmatrix}
  0 & e^{i k\omega m \delta t/2} \\
  e^{-i k\omega m \delta t/2} & 0
\end{pmatrix}
=
\begin{pmatrix}
  e^{i k\omega \delta t/2} & 0 \\
  0 & e^{i k\omega \delta t/2}
\end{pmatrix},
\eeq
which are crucially independent of $m$.
Thanks to this simplification, the matrix representation \eqref{Ut_SL2_2x2} of the evolution operator can be recast as
\beq
\label{Ut_SL2_2x2_ver2}
U(t) \big|_{2\times2}
= e^{i \frac{\pi c}{12L}k^2 t}
  \lim_{M\to\infty}
  \begin{pmatrix}
  0 & e^{i k\omega t/2} \\
  e^{-i k\omega t/2} & 0
\end{pmatrix} \!
  \mathsf{A}^M
  \exp \biggl[ \frac{\pi k}{L} \delta t \begin{pmatrix}
    -i \cosh(2\lambda) & \sinh(2\lambda) \\
    \sinh(2\lambda) & i \cosh(2\lambda)
  \end{pmatrix} \biggr] \!
  \begin{pmatrix} 0 & 1 \\ 1 & 0 \end{pmatrix},
\eeq
where
\beq
\mathsf{A}
=
\exp \biggl[ \frac{\pi k}{L} \delta t \begin{pmatrix}
      -i \cosh(2\lambda) & \sinh(2\lambda) \\
      \sinh(2\lambda) & i \cosh(2\lambda)
  \end{pmatrix} \biggr]
  \begin{pmatrix} e^{i k\omega \delta t/2} & 0 \\ 0 & - e^{i k\omega \delta t/2} \end{pmatrix}
=
\begin{pmatrix} a & b \\ \bar{b} & \bar{a} \end{pmatrix}
  + O(\delta t^2)
\eeq
with $a = \bigl[ 1 - i(\pi k/L) \delta t \cosh(2\lambda) \bigr] e^{i k\omega  \delta t/2}$ and $b = (\pi k/L) \delta t \sinh(2\lambda) e^{-i k\omega  \delta t/2}$.
One can now diagonalize $\mathsf{A}$, noting that all $O(\delta t^2)$ contributions give rise to subleading corrections in $1/M$, to find
\begin{align}
\label{Ut_SL2_2x2_ver3}
U(t) \big|_{2\times2}
& = e^{i \frac{\pi c}{12 L} k^2 t}
  \begin{pmatrix}
    e^{i k\omega t/2} & 0 \\
    0 & e^{-i k\omega t/2}
  \end{pmatrix} \\
& \quad \times
  \begin{pmatrix}
    \cos(k\Omega t/2) + i \frac{\cosh(2\lambda) - L/T}{L\Omega / 2\pi} \sin(k\Omega t/2)
    & \frac{\sinh(2\lambda)}{L\Omega / 2\pi} \sin(k\Omega t/2) \\
    \frac{\sinh(2\lambda)}{L\Omega / 2\pi} \sin(k\Omega t/2)
    & \cos(k\Omega t/2) - i \frac{\cosh(2\lambda) - L/T}{L\Omega / 2\pi} \sin(k\Omega t/2)
  \end{pmatrix}, \nonumber
\end{align}
where we used $\lim_{M\to\infty} (1 + \kappa/M)^{\mu M} = e^{\mu\kappa}$ for $\kappa, \mu \in \RR$.
\end{widetext}

Lastly, we use the ansatz $e^{i[G + \bar{F}L_{-k} + E L_{0} + F L_{k}]} \big|_{2\times2}$ ($G, E \in \RR$ and $F \in \CC$) for the second matrix in Eq.~\eqref{Ut_SL2_2x2_ver3}.
The fundamental representation \eqref{2x2_rep} then allows one to show that Eq.~\eqref{Ut_SL2_Ls} holds for the evolution operator.
This can also be used to find the diffeomorphism $g_t\in\Diff(S^1)$ corresponding to the two last exponentials in Eq.~\eqref{Ut_SL2_Ls}.
Owing to the properties of one-parameter flows, one can indeed show that $U(t) = e^{i\alpha_t} \cU(g_t) \cU(g_0)^{-1}$ in terms of a diffeomorphism $g_t$ as in Eq.~\eqref{gt}, with $\Omega$ given by \eqref{Omega_SL2} and $\zeta(\phii)$ a transformation \eqref{s435} with $\alpha$, $\beta$ as in Eq.~\eqref{alphabeta_SL2_zeta}.

\paragraph{Return amplitudes.}
When the initial condition is $g_0 = f_0$ so that $|\psi(0)\rangle = \cU(f_0) |h\rangle$,
one also needs to express $\cU(f_0)$ as a $2 \times 2$ matrix for $f_{0}$ of the $\mathrm{SL}(2,\RR)$ form \eqref{s435} with $\alpha = \cosh(\lambda)$, $\beta = \sinh(\lambda)$.
This $f_0$ is the one-parameter flow generated by $-(2/k)\sin(k\phii)\partial_{\phii}$ with respect to $\lambda$, so one finds
\beq
\label{cU_f0_2x2}
\begin{aligned}
\cU(f_0) \big|_{2\times2}
& = \exp \biggl( \frac{\lambda}{k} [L_{-k} - L_{k}] \biggr) \big|_{2\times2} \\
& = \begin{pmatrix}
      \cosh(\lambda) & i \sinh(\lambda) \\
      -i \sinh(\lambda) & \cosh(\lambda) \\
    \end{pmatrix}.
\end{aligned}
\eeq
Replicating the steps that led to Eq.~\eqref{Ut_SL2_Ls}, the above can be used to factorize the product of $\cU(f_0)$ with the evolution operator $U(t)$ as
\beq
\label{Ug_t_SL2}
U(t) \cU(f_0)
= e^{i \rho_t} e^{ \chi_t L_{-k} - \bar{\chi}_t L_{k} } e^{i \sigma_t L_0}.
\eeq
Here, $\rho_t$ and $\sigma_t$ are real coefficients given by
\begin{align}
\rho_t
& = \frac{\pi c}{12 L} \Bigl[ k^2 t + \frac{L \sigma_t}{2\pi} \bigl( k^2 - 1 \bigr) \Bigr], \\
\sigma_t
& = - \omega t - \frac{2}{k} \arctan \Bigl( \Bigl[ \frac{2\pi}{L \Omega} - \frac{\omega}{\Omega} \Bigr] \tan(k \Omega t/2) \Bigr), \nonumber
\end{align}
and $\chi_t$ is determined by $\eta_t$ in Eq.~\eqref{eta_t_SL2}, with $\Omega$ given in Eq.~\eqref{Omega_SL2}.
The ray of the vector $U(t) \cU(f_0)|h\rangle$ is only sensitive to $\chi_t,\bar\chi_t$ as in Eq.~\eqref{psi_t_factorized}, since $\rho_t$, $\sigma_t$ only contribute an overall phase $\rho_t + h\sigma_t$ due to Eq.~\eqref{e217}.

Lastly, Eqs.~\eqref{Ut_SL2_2x2_ver3} and~\eqref{cU_f0_2x2} can be combined to represent $\cU(f_0)^{-1} U(t) \cU(f_0)$ as an $\mathrm{SU}(1,1)$ matrix:
\beq
\cU(f_0)^{-1} U(t) \cU(f_0) \big|_{2\times2}
= \begin{pmatrix}
    u_t & w_t \\
    \bar{w_t} & \bar{u_t}
  \end{pmatrix}
\eeq
with $u_t$ as in Eq.~\eqref{eq:u_t} and
\begin{align}
w_t
= \sinh(2\lambda) \biggl[ & \frac{\omega}{\Omega} \sin(k\Omega t/2) \cos(k\omega t/2) \\
& \begin{aligned}
  & - \cos(k\Omega t/2) \sin(k\omega t/2) \\
  & + i \frac{2\pi}{L\Omega} \sin(k\Omega t/2) \sin(k\omega t/2) \biggr].
  \end{aligned} \nonumber
\end{align}
This must, on general grounds, be the nonunitary $2 \times 2$-matrix representation of an element of the form $e^{D} e^{AL_{-k}} e^{BL_{0}} e^{CL_{k}}$.
One finds that
\beq
\begin{gathered}
A = \frac{i}{k} \frac{\bar{w}}{u},
\quad\;
B = -\frac{2}{k} \log(u),
\quad\;
C = \frac{i}{k} \frac{w}{u}, \\
e^{D}
= e^{i \frac{\pi c}{12 L}k^2 t} e^{B \frac{c}{24}(k^2 - 1)}.
\end{gathered}
\eeq
It follows using Eq.~\eqref{e217} that
\beq
\begin{aligned}
\langle \psi(0)| \psi(t) \rangle
& = \langle h| \cU(f_0)^{-1} U(t) \cU(f_0) |h\rangle \\
& = e^D \langle h| e^{BL_0} |h\rangle
  = e^{D + Bh},
\end{aligned}
\eeq
which yields Eq.~\eqref{p0pt_SL2} for the return amplitude.

%==========================================================
\section{Lattice calculations}
\label{App:LattCalcs}
%==========================================================

Let us finally provide details on the numerics used throughout this work.

\paragraph{Lattice Hamiltonian and adjacency matrix.}
We consider a driven inhomogeneous free-fermion chain at half filling.
The operator algebra is generated by fermionic creation (annihilation) operators $c_{j}^{\dagger}$ ($c_{j}$) at sites $j = 1, \ldots, N$, which satisfy the usual anticommutation relations $\{ c_{j\ppr}\pdag, c_{j'}^{\dagger} \} = \delta_{j,j'}$ and $\{ c_{j\ppr}\pdag, c_{j'}\pdag \} = \{ c_{j\ppr}^{\dagger}, c_{j'}^{\dagger} \} = 0$.
The Hamiltonian is
\beq
H_{N}(t)
= 
- \frac{1}{2} \sum_{j=1}^{N} J_{j}\pdag(t) \Bigl( c_{j}^{\dagger}c_{j+1}\pdag + c_{j+1}^{\dagger}c_{j}\pdag \Bigr)
\label{eq:latticemodelnum}
\eeq
with site- and time-dependent nearest-neighbor hoppings
$J_{j}(t) = J_{j}(t+T) > 0$ and periodic boundary conditions, $c_{N+1}^{(\dagger)} = c_{1}^{(\dagger)}$ and $J_{N+1}\pdag(t) = J_{1}\pdag(t)$.
A Jordan-Wigner transformation maps this model on the inhomogeneous spin chain \eqref{eq:H_XXZ_t} with $\Delta = 0$ (up to boundary terms) \cite{Moosavi:PhD:2018}.
We assume that the system size $L$ is much larger than the lattice spacing $L/N$, \ie $N \gg 1$, and that $J_{j}(t)$ vary on mesoscopic length scales.
Then, the model \eqref{eq:latticemodelnum} can be effectively described in the low-energy regime by an inhomogeneous CFT \eqref{fullh} with $c = \bar{c} = 1$, involving both right- and left-moving (chiral and antichiral) components \cite{Moosavi:iCFT:2024}.

The Hamiltonian for any periodic fermionic chain with nearest-neighbor hoppings
can be represented by a periodic Jacobi matrix.
There is no external field in our case, so the matrix is off-diagonal and reads
\beq
\label{Jt_matrix}
\mathsf{J}(t)
\equiv
\begin{pmatrix}
0 & J_{1}(t) & 0 & \ldots & 0 & J_{N}(t)\\
J_{1}(t) & 0 & J_{2}(t) & \ldots & 0 & 0 \\
0 & J_{2}(t) & 0 & \ldots & 0 & 0 \\
\vdots & \vdots & \vdots & \ddots & \vdots & \vdots \\
0 & 0 & 0 & \ldots & 0 & J_{N-1}(t) \\
J_{N}(t) & 0 & 0 & \ldots & J_{N-1}(t) & 0 \\
\end{pmatrix},
\eeq
which is nothing but the adjacency matrix corresponding to the Hamiltonian \eqref{eq:latticemodelnum}.
The latter can be
recast as $H_{N}(t) = (c_1^{\dagger}, c_2^\dagger, \ldots, c_N^{\dagger}) \mathsf{H}(t) (c_1\pdag, c_2\pdag, \ldots c_N\pdag)^T$ with $\mathsf{H}(t) \equiv - \mathsf{J}(t)/2$ interpreted as a single-particle Hamiltonian.
If $\mathsf{U}$ denotes the unitary $N\times N$ matrix that diagonalizes the initial matrix Hamiltonian $\mathsf{H}(0)$, then one can write
\beq
H_{N}(0)
= \sum_{j=1}^{N} E_{j}(0)\gamma_{j}^{\dagger} \gamma_{j}\pdag,
\quad\;
\gamma_{j}
\equiv
\sum_{\ell=1}^{N} (\mathsf{U}^{\dagger})_{j\ell} c_{\ell},
\label{eqe4}
\eeq
where $E_{j}(0)$ denotes the eigenenergies of $\mathsf{H}(0)$ and $\gamma_{j}^\dagger$ ($\gamma_{j}\pdag$) are new fermionic creation (annihilation) operators.
General properties of periodic Jacobi matrices \cite{Moerbeke:1976} imply that the $E_{j}(0)$ can be ordered to be (nonstrictly) increasing in $j = 1, \ldots, N$, with each eigenenergy equal to at most one other.

The many-body ground state is the filled Dirac sea, which corresponds to filling all single-particle states with negative energies.
For simplicity, let $N$ be even.
Assume furthermore that $N/2$ is odd, since that excludes that zero is an eigenenergy \cite{Moerbeke:1976}.
Then, at half filling, all states with $j = 1, \ldots, N/2$ have negative energies,\footnote{For $N$ even, let $S = \diag(1, -1, 1, -1, \ldots, 1, -1)$. Since $\mathsf{J}(t)$ in Eq.~\eqref{Jt_matrix} satisfies $\mathsf{S} \mathsf{J}(t) = - \mathsf{J}(t) \mathsf{S}$, if $\bx$ is an eigenvector of $\mathsf{H}(t) = -\mathsf{J}(t)/2$ with eigenvalue $E$, then $\mathsf{S} \bx$ is also an eigenvector with eigenvalue $-E$.} \ie the ground state is
\beq
|\psi_{N}(0)\rangle 
=
\gamma_{1}^{\dagger} \cdots \gamma_{N/2}^{\dagger}|\mathrm{vac}\rangle,
\label{stst}
\eeq
where $|\mathrm{vac}\rangle$ denotes the state with no fermions.
[For $N$ odd, one must replace $N/2$ by the maximal value $N_\text{max}$ for which $E_{N_\text{max}}(0)$ is negative.] 
Note that general filling is obtained by including the term $- \sum_{j=1}^{N} \mu_{j}(t) (c_j^{\dagger} c_{j}\pdag - 1/2)$ in the Hamiltonian \eqref{eq:latticemodelnum}, with a time-dependent inhomogeneous chemical potential $\mu_j(t)$.
One then needs to fill its corresponding negative-energy one-particle states.
These are states of the one-particle Hamiltonian given by Eq.~\eqref{Jt_matrix}, now with nonzero diagonal elements $\mu_{1}(t), \ldots, \mu_{N}$(t), which must have the same mesoscopic spatial dependence as $J_j(t)$ for there to be an emergent CFT at low energies (see Ch.~5 in \cite{Moosavi:PhD:2018}).

\paragraph{Return probabilities.}
Our considerations involve only fermions at zero temperature, so the (nonchiral) CFT model \eqref{fullh} is expected to be a good low-energy description of the lattice dynamics.
Concretely, we let $|\psi_{N}(0)\rangle$ be the ground state \eqref{stst} of $H_{N}(0)$, which in CFT language corresponds to the vacuum $|h\rangle\otimes|\bar h\rangle = |0\rangle\otimes|0\rangle$ as defined below Eq.~\eqref{e217}.
We then calculate the return probability after one period $T$,
\beq
\label{cL_T_latt}
\begin{gathered}
\Bigl| \langle\psi_{N}(0)|\psi_{N}(T)\rangle \Bigr|^2
=
\Bigl| \langle\psi_{N}(0)| U_{N}(T) |\psi_{N}(0)\rangle \Bigr|^2, \\
U_{N}(T)
\equiv
\overset{\longleftarrow}{\cT}
\exp\!\Bigl(
 -i\int_0^T \dd t\, H_{N}(t)
\Bigr).
\end{gathered}
\eeq
In terms of the creation and annihilation operators introduced in Eq.~\eqref{eqe4}, this can be recast as
\begin{align}
& \Bigl| \langle\psi_{N}(0)|\psi_{N}(T)\rangle \Bigr|^2 \nonumber \\
& =
\Bigl| \langle\mathrm{vac}| \gamma_{1}\pdag \ldots \gamma_{N/2}\pdag U_{N}(T) \gamma_{N/2}^{\dagger} \ldots \gamma_{1}^{\dagger} |\mathrm{vac}\rangle \Bigr|^2 \nonumber \\
& = \Bigl| \langle\mathrm{vac}| [U_{N}(T)^{\dagger} \gamma_{1}\pdag U_{N}(T)] \ldots [U_{N}(T)^{\dagger} \gamma_{N/2}\pdag U_{N}(T)] \nonumber \\
& \quad\; \times \gamma_{N/2}^{\dagger} \ldots \gamma_{1}^{\dagger} |\mathrm{vac}\rangle \Bigr|^2,
\label{latticeloschtechnical}
\end{align}
since $U_{N}(T)$ acting on $\langle\mathrm{vac}|$ only gives a phase.
To compute each factor $U_{N}(T)^{\dagger} \gamma_{j} U_{N}(T)$ here, one needs to discretize the continuous time evolution.
This is done as in Eq.~\eqref{Ut_Trotter} by approximating the evolution under $H_{N}(t)$ as consisting of $M \gg 1$ time steps, each evolved under a static Hamiltonian $H_{N}(m\delta t)$ for a time $\delta t = T/M$.
In other words,
\beq
U_{N}(T)
\approx \prod_{m=0}^{M} e^{-i H_{N}(m\delta t) \delta t},
\eeq
where the approximation improves as $M$ increases.

Let $\mathsf{V}_m$ be the unitary $N\times N$ matrix that diagonalizes the single-particle matrix Hamiltonian $\mathsf{H}(m\delta t)$ corresponding to $H_{N}(m\delta t)$.
Let also $E_{j}(m\delta t)$ denote the eigenenergies of $\mathsf{H}(m\delta t)$.
Defining $\mathsf{W}_m(\delta t) \equiv \mathsf{U}^{\dagger} \mathsf{V}_m\pdag \diag \bigl( e^{-iE_1(m\delta t)\delta t}, \ldots, e^{-iE_N(m\delta t)\delta t} \bigr) \mathsf{V}_m^{\dagger} \mathsf{U}$, it follows that 
\beq
e^{ i H_{N}(m\delta t) \delta t } \gamma_{j} e^{ -i H_{N}(m\delta t) \delta t }
=
\sum_{\ell=1}^{N} \Bigr( \mathsf{W}_m(\delta t) \Bigr)_{\!j\ell\,} \gamma_{\ell}.
\eeq
We use this decomposition for each time step in Eq.~\eqref{latticeloschtechnical}, which reduces the whole calculation to evaluating a correlator of the $\gamma_{\ell}$\hspace{0.5pt}s.
The return probability \eqref{latticeloschtechnical} can finally be approximated as
\beq
\label{psiN0_psiNT_approx}
\Bigl| \langle\psi_{N}(0) |\psi_{N}(T)\rangle \Bigr|^2
\approx \left|\det_{1 \leq i,\ell \leq N/2} \left( \prod_{m=0}^{M} \mathsf{W}_m(\delta t) \right)_{j\ell}\right|^2
\eeq
with the help of Wick's theorem.
This is the approximate formula used to plot the numerical lattice results in Figs.~\ref{introfig_loschmidt} and~\ref{Fig:Numerics}.

To relate the above to our CFT results, let $J_{j}(t) = [J_{t}(x_{j}) + J_{t}(x_{j+1})]/2$
be given by a slowly varying smooth function $J_{t}(x)$ of $x$, for each time $t$.
The hopping (or coupling) between sites $j$ and $j+1$ is thus attributed to the sites themselves.
The CFT propagation velocity, emerging when linearizing the dispersion relation around its two Fermi points, can then be expressed as
$v_{t}(x_j) = J_{t}(x_j) (L/N) \sin(k_F L/N)$, where $k_F$ is the Fermi momentum and $L/N$ is the lattice spacing (\cf Ch.~5 in \cite{Moosavi:PhD:2018}).
Exactly at half filling, these satisfy $k_F L/N = \pi/2$, in which case $v_{t}(x_j) = J_{t}(x_j) L/N$, as used in Figs.~\ref{introfig_loschmidt} and~\ref{Fig:Numerics}.
That said, the same procedure is equally valid away from (but sufficiently near) half filling.

%==========================================================


%==========================================================
\end{document}